\documentstyle[aps,multicol,psfig,eqsecnum]{revtex}  
\input epsf
\begin{document}
\def\st{\scriptstyle}
\def\bein{{\it bein\/}}
\def\beine{{\it beine\/}}
\def\half{{\scriptstyle{\scriptstyle 1\over{\scriptstyle 2}}}}
\def\haim{{\scriptstyle{\scriptstyle i\over{\scriptstyle 2}}}}
\def\fop{{\Omega}}
\def\pop{{W}}		
\def\iop{{w}}		
\def\sst{\scriptstyle}
\def\vtripL#1{{}^{\alpha_{#1}}_{\ell_{#1} m_{#1}}}
\def\vtripW#1{{}^{\alpha_{#1}}_{1         m_{#1}}}
\def\hk{{\hat{\bf k}}}
\def\hl{{\hat{\ell}}}
\def\hm{{\hat{m}}}
\def\hz{{\hat{0}}}
\def\bfhz{\hat{\bf 0}}
\def\hy{{\hat{Y}}}
\def\hs{{\hat{\bf s}}}
\def\cp{{\epsilon}}	
\def\sce{{SCE}}		
\def\co#1{{C_{#1}}}	
\def\sck#1{{k_{m_{#1}}^{\alpha_{#1}\,\ast}}}
\def\sckW#1{{k_{1\,m_{#1}}^{\alpha_{#1}\,\ast}}}
\def\bfpc#1{{{\cal C}^{(#1)}}}
\def\liso{\left \{         }	
\def\riso{\right\}_{\delta}}	
\def\poly{{\cal P}}
\def\tens{{\cal T}}
\def\con{{\cal C}}
\def\aaop{\Sigma}
\def\pacf{\Gamma}
\def\bfmu{\bbox{\mu}}
\def\piconv{\kappa}
\draft
\title{Statistical mechanics of permanent random atomic and 
molecular networks: \\
Structure and heterogeneity of the amorphous solid state}
\author{Konstantin A.~Shakhnovich and Paul M.~Goldbart}
\address{Department of Physics, 
	University of Illinois at Urbana-Champaign, \\
	1110 West Green Street, Urbana, Illinois 61801-3080, U.S.A.} 
 \date{January 18, 1999}
\maketitle
\begin{abstract}
Under sufficient permanent random covalent bonding, a fluid of atoms or 
small molecules is transformed into an amorphous solid network.  Being 
amorphous, local structural properties in such networks vary across the 
sample.  A natural order parameter, resulting from a 
statistical-mechanical approach, captures information concerning this 
heterogeneity via a certain joint probability distribution.  This joint 
probability distribution describes the variations in the positional and 
orientational localization of the particles, reflecting the random 
environments experienced by them, as well as further information 
characterizing the thermal motion of particles.  A  complete solution, 
valid in the vicinity of the amorphous solidification transition, is 
constructed essentially analytically for the amorphous solid order 
parameter, in the context of the random network model and approach 
introduced by Goldbart and Zippelius 
[{\sl Europhys.\ Lett.\/}~{\bf 27\/}, 599 (1994)].  
Knowledge of this order parameter allows 
us to draw certain conclusions about the stucture and heterogeneity of 
randomly covalently bonded atomic or molecular network solids in the 
vicinity of the amorphous solidification transition.  {\it Inter alia\/}, 
the positional aspects of particle localization are established to have 
precisely the structure obtained perviously in the context of vulcanized 
media, and results are found for the analogue of the spin glass order 
parameter describing the orientational freezing of the bonds between 
particles.
\end{abstract}
%
%
\section{Introduction and overview}
\label{SEC:intro}
The purpose of this Paper is to address the statistical structure of the
amorphous solid state via a simple model of a three-dimensional vitreous
medium consisting of covalently bonded atoms (or low-molecular-weight
molecules)~\cite{REF:Zallen}. We shall do so essentially analytically by
making use of techniques drawn from the field of the statistical
mechanics of systems having quenched randomness~\cite{REF:MPVbook}.  The
model of vitreous media on which we shall focus is that introduced by
Goldbart and Zippelius~\cite{REF:GZeplSG}, which takes as ingredients a
thermodynamically large number of particles between which some large
number of permanent random covalent bonds are introduced.  The quenched
randomness is encoded in the information describing which pairs of
particles are covalently bonded; the remaining (annealed) degrees of
freedom correspond to the unconstrained positions of the particles and
the orientations of the orbitals. This model exhibits a continuous
equilibrium phase transition from the liquid state to the amorphous
solid state when the density of introduced bonds exceeds a certain
critical value. It is on the structure and heterogeneity of this state
that we hope to shed some light.

As an example of the type of medium we have in mind, consider networks
formed by the polycondensation of ${\rm Si(OH)_{4}\/}$ molecules, during
which ${\rm H_{2}O\/}$ is eliminated between pairs of hydroxyl (OH)
groups on certain randomly selected pairs of ${\rm Si(OH)_{4}\/}$
molecules so as to form Si--O--Si bonds.  The amorphous solidification
of such media has been studied in many experiments; we cite as an
example those of Gauthier-Manuel et al.~\cite{REF:Gauthier}.  As it is
our intention to develop a rather general model of random networks, and to 
focus on universal properties, it is not necessary for us to incorporate
the specific details of the medium.  For example, we shall not be
accounting for the bond geometry associated with the so-called bridging
oxygen atoms between the silicon atoms.  In the model, both types of Si
orbitals, those connected to hydroxyl groups and those connected to
bridging oxygen atoms, will simply be referred to as 
\lq\lq orbitals\rlap.\rq\rq\  A second example of the type of media we
have in mind is provided by amorphous silicon
networks~\cite{REF:Zallen}, especially those in which hydrogen
passivates bonds unconnected to other silicon atoms.

The structural characterization of the vitreous state that we shall
construct will be statistical in nature, reflecting the intrinsic
heterogeneity of the environments that the constituent particles in
vitreous media inhabit.  It will take the form of a {\it joint
probability distribution} characterizing the fraction of particles that
are localized in the vitreous state, and will describe the spatial
extent of the thermal fluctuations in their positions, the degree and
character of the thermal fluctuations in the orientations of the
orbitals that are capable of participating in covalent bonds, and the
strength and nature of the correlations between the thermal fluctuations
in the particle positions and the orbital orientations.  Moreover, rather
than dealing with media having a specific architecture (i.e., a specific
realization of introduced bonds), we shall consider an ensemble of
architectures, all characterized by a common parameter governing the
probability that a permanent chemical bond was formed between any pair
of nearby orbitals.

A statistical description of an amorphous solid state has previously
been developed and explored in the context of vulcanized (i.e., randomly
permanently crosslinked) {\it macromolecular}
media~\cite{REF:cast,REF:review}.  This description, which addresses the
distribution of spatial extents of thermal position-fluctuations (i.e.,
localization lengths) has been confirmed by computer
simulations~\cite{REF:SJB_MP}, and rather general, model-nonspecific
arguments in favor of the broad applicability of the
description have also been presented~\cite{REF:Peng}.  For
any particular version of random media (e.g. the macromolecular media of 
Refs~\cite{REF:cast,REF:review} or the vitreous media considered in the
present paper), what determines the specific content of the
statistical description of the amorphous solid state is the form of the
random constraints that the permanent covalent bonding imposes, and the
resultant form taken by the amorphous solid order parameter.  In the present
context of vitreous media, the constraints, as we shall see below, are
more intricate than they are in the macromolecular vulcanization context and,
accordingly, the order parameter is more intricate and the statistical
content more elaborate: it accounts not only for the heterogeneity (i.e.,
the distribution over the sample) of positional localization lengths but
also for the distribution of orbital-orientation thermal fluctuations,
the strengths of the position-angle thermal fluctuation correlations,
and the statistical correlations between these physical 
characteristics~\cite{REF:qualif}.

In Ref.~\cite{REF:GZeplSG}, in addition to introducing the model of random
network forming media considered here, and formulating the question of
the phase transition to (and structure of) the amorphous solid state via
statistical-mechanical techniques, Goldbart and Zippelius made a simple
variational mean-field theory for the amorphous solid state in which all
particles shared a common localization length and all orbitals shared a
common extent of their angular localization. The positional and angular
localization parameters were then solved for, self-consistently, and it
was found that, at a certain critical value of the density of formed
bonds, a continuous transition to an amorphous solid state occurs,
beyond which the inverse of the positional localization parameter grows
continuously from zero. It was also found that, in response to the onset
of positional localization, orientational localization of the orbitals 
sets in. Owing to the restricted form of the variational hypothesis for 
the order parameter adopted in Ref.~\cite{REF:GZeplSG}, specifically that 
it did not allow for the possibility that only a fraction of the particles 
would become localized at the transition, the critical bond density was 
over-estimated in Ref.~\cite{REF:GZeplSG}.  (The correct critical 
density was, however, known from the linear stability analysis of the fluid 
state.)\thinspace\ Later work by Theissen et al.~\cite{REF:alloys}, 
in addition to allowing for networks comprising particles of various 
valencies, cured the difficulty of the critical bond density, by 
broadening the variational hypothesis to allow for a localized fraction 
(although it still only allowed for a single value for the positional 
and orientation localization parameters, and did not account for 
correlations between the thermal fluctuations of positions and 
orientations).  

What, then, is the nature of the amorphous solid state?  If the number of
permanent random covalent bonds introduced between particles is smaller
than a certain critical value then the effect of these bonds is to bind
at least some of the particles into random permanent molecules of a
variety of types (varying in size and architecture), each of which, given
sufficient time, will wander ergodically through the volume of the
container, i.e., the equilibrium state of the system is fluid.  If,
however, the number of bonds introduced is greater than the critical
value then their effect is to bind a nonzero fraction of the particles
into a {\it macroscopically large disordered molecule} that extends throughout
the container, the remaining fraction of particles remaining
disconnected from the macroscopic molecule and capable of wandering
across the container, given sufficient time. By contrast, the particles that
constitute the extensive molecule will be localized in the vicinity of
random preferred spatial positions, about which their positions will
undergo thermal fluctuations extending only over a limited spatial
regime (which will vary randomly in magnitude from particle to particle,
reflecting the random architecture of the network), and these particles
will confer a rigidity on the entire system, so that the equilibrium
state of the system will no longer be fluid and will, instead, be solid. 
Moreover, the orbitals attached to localized particles will exhibit
most probable orientations, about which they will fluctuate thermally,
the extent and nature of these fluctuations also varying randomly from
orbital to orbital.  In addition, the thermal fluctuations in the positions of
the particles and the orientations of the bonds connecting them will
be correlated, to an extent that varies randomly from particle to particle.
The unconventional nature of this the amorphous solid state is worth
emphasizing: 
(i)~only a fraction of the particles will be localized;
(ii)~the mean positions of the localized particles will be random, as will be  
(iii)~the spatial extent of the positional fluctuations of the particles, 
(iv)~the orientational fluctuations of the orbitals, and
(v)~the correlations between these fluctuations (these parameters being 
characterized by a joint probability distribution); and 
(vi)~there will be no hint of crystallinity beyond the shortest of 
lengthscales (i.e., the bond length), beyond these lengthscales the 
symmetries of the amorphous solid state being those of the liquid state.

Our principal aims are to construct a statistical characterization of
the structure and heterogeneity of the amorphous solid state exhibited by a model of
permanently randomly bonded vitreous media in the vicinity of the
solidification transition, and to provide a physical interpretation of
this characterization. We shall do this by constructing the self-consistency 
equation for the amorphous solid order parameter, valid in the 
vicinity of the solidification transition, and obtaining an exact solution of 
this self-consistent equation.
  
This Paper is organized as follows. In Secs.~\ref{SEC:elements}
and~\ref{SEC:OPdef} we shall proceed kinematically, describing the
model that we shall be considering, and analyzing a suitable order
parameter defined in terms of the positions of the constituent particles and the
orientations of their orbitals. Continuing kinematically, we shall
explore the structure of this order parameter, and elucidate the
physical information that it encodes. Then, in Secs.~\ref{SEC:RSM}
and~\ref{SEC:MFAsetup}, we shall address the model, regarding the formed
bonds as quenched random variables that vary from realization to
realization.  By using equilibrium statistical mechanics, invoking
the replica technique to deal with the quenched randomness, and making a
mean-field hypothesis, we shall develop a self-consistent equation for
the order parameter.  By making a natural physical hypothesis for the
form of the solution we shall, in Sec.~\ref{SEC:SCEsol}, solve exactly for this
order parameter in the regime in which the thermal fluctuations of the
particle positions and orbital orientations are strong (i.e., near the
solidification transition). Finally, in Sec.~\ref{SEC:PhysInfo} we shall
extract from our solution a wide array of physical diagnostics
characterizing the amorphous solid state and, in Sec.~\ref{SEC:summary}
we shall make some concluding remarks. 
We emphasize that throughout this work we shall be proceeding
analytically, except that we shall make use of the numerically-obtained
scaling function (of a single variable) central to the characterization
of vulcanized macromolecular matter described in
Refs.~\cite{REF:cast,REF:review}.
\section{Elements of the model}
\label{SEC:elements}
\begin{figure}[hbt]
 \epsfxsize=4.0in
 \centerline{\epsfbox{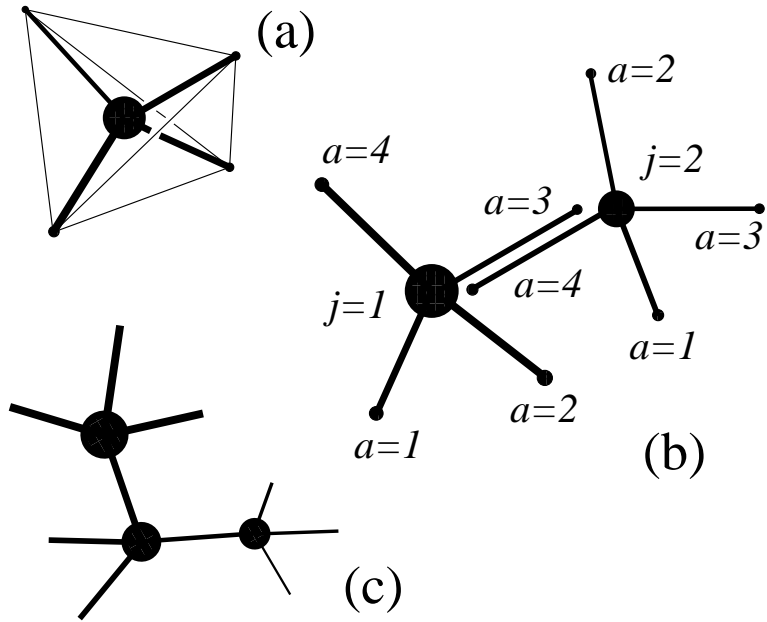}}
\vskip0.5truecm
\caption{Particle structure, bond, and network formation.
(a)~A single particle with the near-tetrahedral equilibrium
structure of its orbitals.  In this example the 
number of orbitals per atom $A$ is $4$ (as would be the case, 
e.g., for networks of Si atoms).
(b)~Formation of a covalent bond between two particles 
(the participating orbitals are slightly separated, for clarity). 
(c)~A collection of three particles bonded together, 
forming the beginnings of a random network.}
\label{FIG:legs_str}
\end{figure}

The model of vitreous media which we shall focus on is that introduced
in Ref.~\cite{REF:GZeplSG}, which takes as its ingredients a
thermodynamically large number $N$ of particles, moving in a large
three-dimensional cube of volume $V$ (on which we impose periodic boundary 
conditions), at least some of which particles are permanently randomly 
bonded together to form a
random network.  At the kinematic level, the particles 
(labeled by $j=1,\ldots,N$)
are characterized by their position vectors $\{{\bf c}_{j}\}_{j=1}^{N}$,
along with the $NA$ unit vectors 
$\{{\bf s}_{j,a}\}_{a=1}^{A}{}_{j=1}^{N}$ 
describing the spatial orientations of
the $A$ orbitals that radiate from each of the particles $j$.
Note that we shall be measuring lengths in units such that 
orbitals have length unity. Figure~\ref{FIG:legs_str} illustrates the
structure of the particles and the formation of a continuous random network
out of them.
The orbitals radiating from a given particle tend to repel one another.  
For example, in the absence of any external perturbing forces, 
all things being equal, the orbitals of a 
four-orbital particle would point towards the vertices of a regular 
tetrahedron, as shown in Fig.~\ref{FIG:legs_str}(a).  
Rather than give a detailed 
specification of the interactions that embody this orbital-orbital 
repulsion, we shall encode the effects of such interactions into a 
sequence of parameters that characterize the correlations between the 
orientations of the orbitals of a single particle.  For example, we shall 
find ourselves needing the correlator of the orientations of two 
{\it distinct} orbitals ($a_{1}$ and $a_{2}$) of a single particle (in the 
fluid state), say the $j^{\rm th}$: 
\begin{equation}
\Big\langle
Y_{\ell_{1}m_{1}}^{\ast}({\bf s}_{j,a_{1}})\,
Y_{\ell_{2}m_{2}}       ({\bf s}_{j,a_{2}})
\Big\rangle_{1,1}
={1\over{4\pi}}\,
 \delta_{\ell_{1},\ell_{2}}\,
 \delta_{m_{1},m_{2}}\,\co{\ell}, 
\label{EQ:corrdef1}
\end{equation}
which we have parametrized in terms of the real numbers
$\{\co{\ell}\}_{\ell=0}^{\infty}$ (with $\co{0}\equiv 1$) that reflect
the extent to which the orbitals interact~\cite{REF:corrconst}.  The angle 
brackets $\langle\cdots\rangle_{1,1}$, which we discuss below in 
Sec.~\ref{SEC:PartFunc}, denote thermal averaging with respect to a the 
single-particle Hamiltonian, which incorporates the intra-particle 
interactions.  The form of this correlator
follows from the isotropy of the distribution of the orbital
orientations in the fluid state. We shall not find ourselves making
explicit use of the correlator of the orientations of {\it three} distinct
orbitals of a single particle.  However, a simple symmetry-dictated form
for it can readily be constructed, by making use of Wigner~3-$j$
technology, if one wishes to compute explicitly components of the order
parameter that depend on it.

Having described the issue of a single particle and its orbitals, we 
now turn to the issue of the permanent random covalent bonds between 
pairs of particles, and how we are to describe them.  
We regard such bonding as introducing constraints on the 
relative location and relative orientation of the particles and 
orbitals participating 
in the bond.  Specifically, we model the situation in which 
particles $j$ and $j'$ are bonded via orbitals $a$ and $a'$ by the 
constraints 
\begin{mathletters}
\begin{eqnarray}
{\bf c}_{j}+{1\over{2}}{\bf s}_{j,a}
&=&
{\bf c}_{j^{\prime}}+{1\over{2}}{\bf s}_{j^{\prime},a^{\prime}},
\\
{\bf s}_{j,a}&=&-{\bf s}_{j^{\prime},a^{\prime}}, 
\end{eqnarray}%
\end{mathletters}%
as shown in Fig.~\ref{FIG:legs_str}(b).  
We denote by the number $M$ and the collection 
$\big\{j_{e},j^{\prime}_{e};a_{e},a^{\prime}_{e}\big\}_{e=1}^{M}$ 
a specific realization of $M$ bonds (i.e. a specific architecture). 

Of course, the particles in the fluid interact with one another, 
regardless of whether or not bonds have been introduced.  We shall
assume that pairwise interactions, depending on the relative separation
and orientation of the orbitals, exist between all particles.  The
crucial consequence that we assume these interactions to have is that
they stabilize the system with respect to the formation of
macroscopically inhomogeneous or anisotropic states, such as regular
crystalline, liquid crystalline, molecular crystalline or globular states.
\section{Amorphous solid order parameter:  
Random positional and orientational localization}
\label{SEC:OPdef}
Following the ideas of Ref.~\cite{REF:GZeplSG}, which represent an
elaboration of ideas discussed in Refs.~\cite{REF:Gold}, we adopt as the
order parameter characterizing the amorphous solid state the entity
\begin{equation}
\left[
{1\over{N}}\sum_{j=1}^{N}
{1\over{A}}\sum_{a=1}^{A}\,
\prod_{\alpha=1}^{n}
\Big\langle
{\rm e}^{-i{\bf k}^{\alpha}\cdot{\bf c}_{j}}\,
Y_{\ell^{\alpha} m^{\alpha}}^{\ast}({\bf s}_{j,a})
\Big\rangle
\right], 
\label{EQ:NZRop}
\end{equation}
where the angle brackets $\langle\cdots\rangle$ (with no subscripts) 
indicate a statistical-mechanical ensemble average over configurations of 
the particles, subject to a given collection of permanent random 
constraints (i.e., bonds), and the square brackets $[\cdots]$ indicate an 
average over realizations of the bonds. This order parameter, which 
involves products of replicas of a single ensemble average, depends on
the collections of wave vectors $\{{\bf k}^{\alpha}\}_{\alpha=1}^{n}$ and
angular momentum indices $\{\ell^{\alpha}, m^{\alpha}\}_{\alpha=1}^{n}$.
 
Let us examine this order parameter, first, in order to ascertain the
nature of the physical states that it is capable of diagnosing, and then
to understand the type of statistical information that it encodes.
\subsection{Detection of random positional and orientational localization}
\label{SEC:detector}
Consider the order parameter given by Eq.~(\ref{EQ:NZRop}), and 
suppose that we elect to set $\ell^{\alpha}=0$ (for $\alpha=1,\ldots,n$).  
Then the order parameter becomes, up to irrelevant factors of $4 \pi$:
\begin{equation}
\left[\,
{1\over{N}}\sum_{j=1}^{N}
\prod_{\alpha=1}^{n}
\Big\langle
\exp(-i{\bf k}^{\alpha}\cdot{\bf c}_{j})\,
\Big\rangle\right].
\label{EQ:NZRPL}
\end{equation}
As discussed in detail, e.g., in Ref.~\cite{REF:review}, and also below 
in Sec.~\ref{SEC:isosector}, these components of the order parameter 
are capable of detecting the 
spontaneous random freezing of particle positions (without regard to 
the behavior of the orbital orientations).  More specifically, 
via the wave vector dependence, these order parameter components yield
information about the fraction of particles that are positionally 
localized, as well as the statistical distribution of their 
positional localization lengths.

Suppose, instead, that we set ${\bf k}^{\alpha}={\bf 0}$ 
(for $\alpha=1,\ldots,n$) in Eq.~(\ref{EQ:NZRop}).  
Then the order parameter becomes
\begin{equation}
\left[
{1\over{N}}\sum_{j=1}^{N}{1\over{A}}\sum_{a=1}^{A}
\prod_{\alpha=1}^{n}
\Big\langle
Y_{\ell^{\alpha} m^{\alpha}}^{\ast}({\bf s}_{j,a})
\Big\rangle\right].
\label{EQ:NZROL}
\end{equation}
As discussed, e.g., in Refs.~\cite{REF:GZeplSG,REF:alloys}, and also
below in Sec.~\ref{SEC:aniso0110}, this component of the order parameter
is capable of detecting the spontaneous random freezing of orbital
orientations (without regard to the behavior of the particle positions).
More specifically, via its dependence on the angular indices
$\{\ell^{\alpha},m^{\alpha}\}_{\alpha=1}^{n}$, 
this order parameter yields information about the extent and character
of the orientational localization of the orbitals.  It is via this
component of the order parameter that the most direct contact is made
with the Edwards-Anderson order parameter for Heisenberg spin glasses, 
which detects the random orientational freezing of magnetic moments.  
For example, choosing 
$\{\ell^{1},\ldots,\ell^{n}\}=\{1,1,0,\ldots,0\}$, and 
contracting appropriately on $m^{1}$ and $m^{2}$ we obtain 
\begin{equation}
\sum_{m^{1},m^{2}=-1}^{1}(-1)^{m^{1}}\,\delta_{m^{1}+m^{2},0}
\left[
{1\over{N}}\sum_{j=1}^{N}{1\over{A}}\sum_{a=1}^{A}
\Big\langle
Y_{1 m^{1}}^{\ast}({\bf s}_{j,a})
\Big\rangle\,
\Big\langle
Y_{1 m^{2}}^{\ast}({\bf s}_{j,a})
\Big\rangle
\right]=
{3\over{4\pi}}
\left[
{1\over{N}}\sum_{j=1}^{N}{1\over{A}}\sum_{a=1}^{A}
\Big\langle{\bf s}_{j,a}\Big\rangle
	\cdot
\Big\langle{\bf s}_{j,a}\Big\rangle
\right], 
\label{EQ:EAexample}
\end{equation}
thus recovering the familiar Edwards-Anderson form.  
More generally, the order parameter for random networks exhibits 
the unconventional features that the index $\ell$ can be greater
than unity (so that higher multipole moments of the distribution of
orientations can be accessed), as well as that a full characterization 
of the orientational freezing requires information from components 
with more than the familiar pair of thermal expectation values.

The third category of information results from examining the components 
of the order parameter corresponding to nonzero values of both 
$\{{\bf k}^{\alpha}\}_{\alpha=1}^{n}$ and 
$\{\ell^{\alpha}\}_{\alpha=1}^{n}$. 
First, consider the subcase for which in every replica $\alpha$ at 
most one of $\ell^{\alpha}$ and ${\bf k}^{\alpha}$ is nonzero.  
An example of such an order parameter component is 
\begin{equation}
\left[
{1\over{N}}\sum_{j=1}^{N}{1\over{A}}\sum_{a=1}^{A}
\Big\langle
\exp(-i{\bf k}^{1}\cdot{\bf c}_{j})
\Big\rangle
\Big\langle
\exp(-i{\bf k}^{2}\cdot{\bf c}_{j})
\Big\rangle
\Big\langle
Y_{\ell^{3} m^{3}}^{\ast}({\bf s}_{j,a})
\Big\rangle
\Big\langle
Y_{\ell^{4} m^{4}}^{\ast}({\bf s}_{j,a})
\Big\rangle
\right].
\label{EQ:NZRscA}
\end{equation}
Such components measure the statistical correlations between the strengths 
of positional and angular localization across the sample.  
{\it Inter alia\/}, such components
address the question: if a certain particle is strongly localized
positionally, how likely are the attached orbitals to be
strongly localized orientationally.

Next, consider the general case in which some replicas $\alpha$ have 
nonzero values for both ${\bf k}^{\alpha}$ and $\ell^{\alpha}$. 
Such components provide information on the extent to which positional and 
orientational {\it thermal} fluctuations are correlated. For example, as is
discussed in more detail below, by setting ${\bf k}^{\alpha} = {\bf 0}$ in all
replicas $\alpha$ except replicas $1$ and $2$, and by also setting 
$\ell^1=\ell^2=1, m^1=m^2=0$, and $\ell{^\alpha}=m^{\alpha}=0$ 
in the remaining replicas $\alpha$, we would
obtain access to the disorder average of the quantity
\begin{equation}
\Big\langle
\big(
{\bf c}_j-\langle{\bf c}_j\rangle
\big)\big(
{\bf s}_{j,a} -\langle {\bf s}_{j,a} \rangle
\big)_z
\Big\rangle
\, \cdot \,
\Big\langle
\big(
{\bf c}_j-\langle{\bf c}_j\rangle
\big)\big(
{\bf s}_{j,a} -\langle {\bf s}_{j,a} \rangle
\big)_z
\Big\rangle,
\end{equation}
which is a direct measure of the extent of the above-mentioned 
position-angle fluctuation correlations. 

Let us pause to emphasize the three levels of randomness presented by 
random network forming media.  There is {\it thermal\/} randomness, 
by which we mean the familiar thermal motion of the particles and 
orbitals.  Then there is {\it architectural\/} randomness, resulting 
from the random manner in which covalent bonds are formed.  Finally, 
there is {\it microstructural\/} randomness, i.e., the 
heterogeneity of the emergent solid state.  This last level of 
randomness we capture statistically in a joint probability 
distribution that characterizes the nature of the thermal motions.
\subsection{Isolating the fraction of positionally localized particles}
\label{SEC:LimitFraction}
The most basic piece of information describing the amorphous solid state 
concerns the value of the fraction $q$ of the $N$
particles that are localized positionally, regardless of the value of
their localization lengths and the angular localization of the orbitals
attached to them.  As shown in Refs.~\cite{REF:GZeplSG,REF:review}, this 
fraction $q$ can be accessed via the order
parameter~(\ref{EQ:NZRop}) in the following way:  set 
$\ell^{\alpha}=m^{\alpha}=0$ for $\alpha = 1,\ldots,n$, and
then pass to the limit $\hk\to{\bfhz}$ via a sequence for which
$\sum_{\alpha=1}^{n}{\bf k}^{\alpha}={\bf 0}$. The resulting quantity
{\it is\/} the fraction $q$.  The reason for this is that whereas the 
value of $\langle \exp \left(-i{\bf k}\cdot{\bf c}_{j} \right) \rangle$ 
at ${\bf k}={\bf 0}$ is
strictly unity, the limiting value of
$\langle \exp \left( -i{\bf k}\cdot{\bf c}_{j} \right) \rangle$ (as
${\bf k}\to{\bf 0}$) is unity for positionally
localized particles, but zero for delocalized particles. 
For the sake of convenience, we shall refer to the localized fraction 
$q$ as the solid fraction, 
and the delocalized fraction $1-q$ as the liquid fraction.
\subsection{Distribution of positional and angular 
localization characteristics}
\label{SEC:locchardist}
To further elucidate the physical information regarding the positional
and orientational localization of the particles and orbitals contained 
in the order parameter, we now construct a physically motivated form for 
the order parameter~(\ref{EQ:NZRop}) in terms of certain {\it localizational 
characteristics\/}---quantities that describe the positional 
and orientational localization of particles and orbitals.
\begin{figure}[hbt]
 \epsfxsize=3.0in
 \centerline{\epsfbox{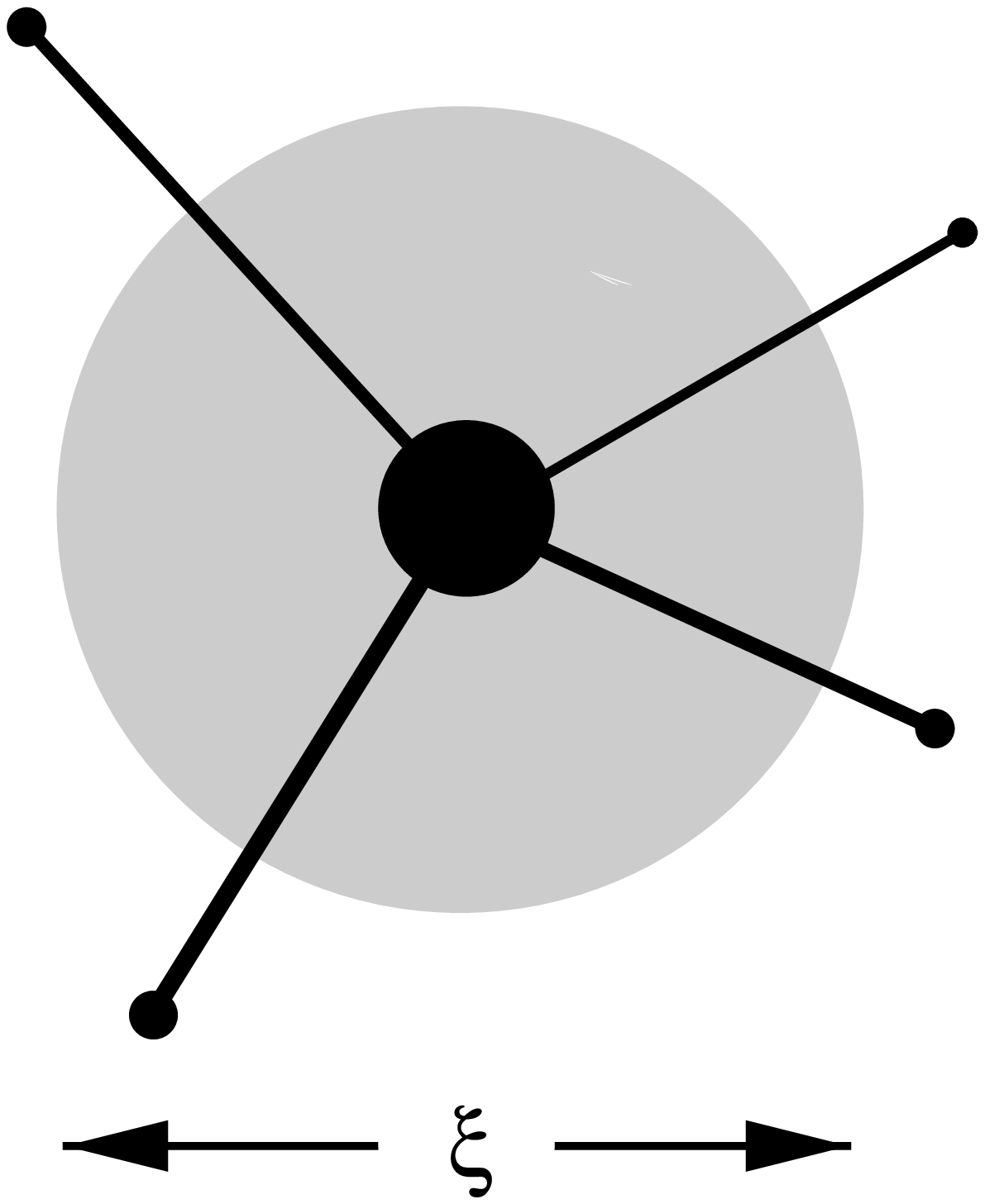}}
\caption{Positional localization of particles.  The characteristic extent 
of the thermal fluctuations of the position of the particle is represented 
by the gray circle and measured by the localization length $\xi$.}
\label{FIG:fluc_pos}
\end{figure}%
We begin by considering the contribution from a single particle $j$ and a 
single orbital $a$ attached to it, i.e., the expectation value
\begin{equation}
\Big\langle{\rm e}^{-i{\bf k}\cdot{\bf c}_{j}}\,
Y_{\ell m}^{\ast}({\bf s}_{j,a})\Big\rangle. 
\label{EQ:onepart}
\end{equation}
This function is the {\it characteristic function\/} of the joint thermal 
probability distribution describing the equilibrium localization of the 
position of the particle and the localization of the orientation of the 
orbital, as well as correlations between the thermal fluctuations between 
this position and orientation.  First, we consider the $(1-q)N$ particles 
in the delocalized fraction.  For such particles we have
$\langle{\rm e}^{-i{\bf k}\cdot{\bf c}_{j}}\,
Y_{\ell m}^{\ast}({\bf s}_{j,a})\rangle=
(4\pi)^{-1/2}\,
\delta_{{\bf k},{\bf 0}}\,
\delta_{{\ell},0}\,
\delta_{m,0}$.
For the remaining $qN$ localized particles we first extract from the 
expectation value~(\ref{EQ:onepart}) the phase factor associated 
with the mean position ${\bfmu}_j \equiv \langle{\bf c}_j\rangle$
of the particle, 
to obtain ${\rm e}^{-i{\bf k}\cdot{\bfmu}_{j}}
\langle{\rm e}^{-i{\bf k}\cdot({\bf c}_{j}-{\bfmu}_{j})}\,
Y_{\ell m}^{\ast}({\bf s}_{j,a})\rangle$, and then express the 
resulting quantity in terms of disconnected and connected pieces: 
\begin{eqnarray}
\Big\langle
{\rm e}^{-i{\bf k}\cdot{\bf c}_{j}}\,
Y_{\ell m}^{\ast}({\bf s}_{j,a})\Big\rangle
&=&
{\rm e}^{-i{\bf k}\cdot{\bfmu}_{j}}
\left\{
\Big\langle{\rm e}^{-i{\bf k}\cdot({\bf c}_{j}-{\bfmu}_{j})}\Big\rangle\,
\Big\langle Y_{\ell m}^{\ast}({\bf s}_{j,a})\Big\rangle
\right.
\nonumber
\\
&&\left.\qquad\qquad
+\,\Big\langle
\Big(
{\rm e}^{-i{\bf k}\cdot({\bf c}_{j}-{\bfmu}_{j})}-
\langle{\rm e}^{-i{\bf k}\cdot({\bf c}_{j}-{\bfmu}_{j})}\rangle
\Big)
\Big(
Y_{\ell m}^{\ast}({\bf s}_{j,a})- 
\langle Y_{\ell m}^{\ast}({\bf s}_{j,a})\rangle
\Big)
\Big\rangle
\right\}.
\label{EQ:corr1}
\end{eqnarray}%
\begin{figure}[hbt]
 \epsfxsize=2.8in
 \centerline{\epsfbox{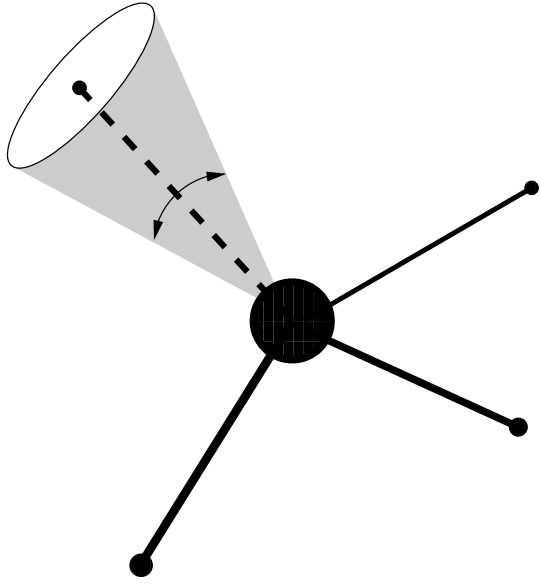}}
\vskip0.5truecm
\caption{Orientational localization of orbitals. The orientation of the 
orbital fluctuates thermally about its most probable value (broken line), 
the characteristic scale of these fluctuations being represented by the 
gray cone.}
\label{FIG:fluc_ori}
\end{figure}%
On the right hand side of this expression, the disconnected (i.e., the first)
piece contains two factors: (i)
$\left\langle\exp\big(-i{\bf k}\cdot({\bf c}_{j}-{\bfmu}_{j})\big)\right\rangle$, 
which describes the positional localization of the particle
(see Fig.~\ref{FIG:fluc_pos}); and (ii) 
$\left\langle Y_{\ell m}^{\ast}({\bf s}_{j,a})\right\rangle$, 
which describes the orientional localization of the orbital 
(see Fig.~\ref{FIG:fluc_ori}).  
If we approximate the first factor in the connected piece by making 
use of the standard cumulant expansion, by
letting $3\xi_{j}^{2}$ denote the (finite) mean square fluctuations 
$\big\langle\left({\bf c}_{j}-\langle{\bf c}_{j}\rangle\right)
\cdot\left({\bf c}_{j}-\langle{\bf c}_{j}\rangle\right)\big\rangle$
in the position of the particle, and by following this strategy to 
second order, then we arrive at the approximation
\begin{equation}
\Big\langle\exp\big(-i{\bf k}\cdot({\bf c}_{j}-{\bfmu}_{j})\big)\Big\rangle
\approx
\exp(-k^{2}\xi_{j}^{2}/2).
\label{EQ:posapprox}
\end{equation}

\begin{figure}[hbt]
 \epsfxsize=2.8in
 \centerline{\epsfbox{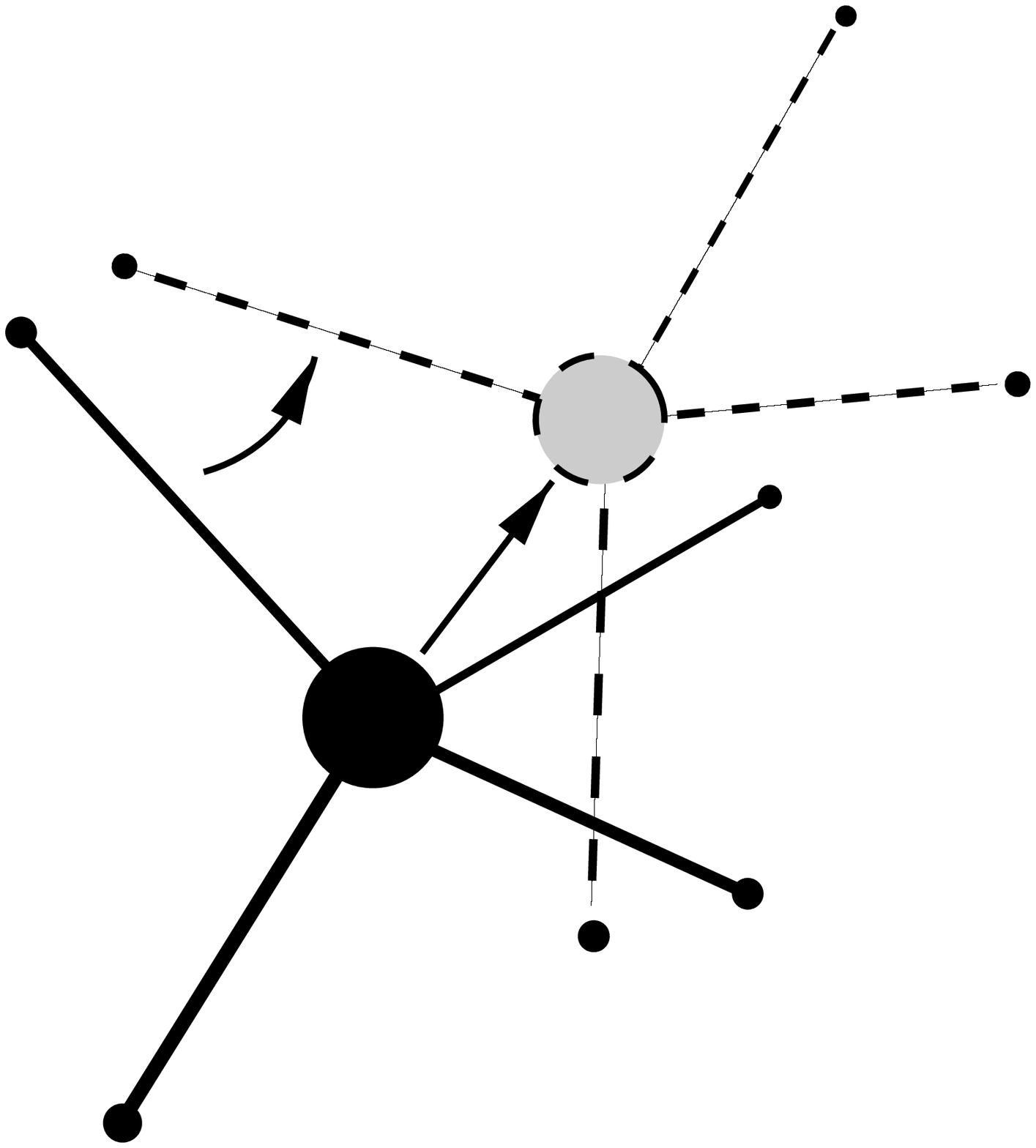}}
\vskip0.5truecm
\caption{Orientational-positional thermal fluctuation correlations. 
Imagine that the particle is connected to a rather immobile 
particle by the top left orbital.  As the particle moves 
to the shaded position its orbitals reorient accordingly.}
\label{FIG:fluc_cor}
\end{figure}%
As for the connected (i.e., the second) term on the right-hand side of
Eq.~(\ref{EQ:corr1}), it describes correlations between the fluctuations 
in the particle position and the orbital orientation 
(see Fig.~\ref{FIG:fluc_cor}).  In the same way 
that we have introduced the diagnostic $\xi$ to characterize positional 
localization, we now introduce two further diagnostics:
\begin{mathletters}
\begin{eqnarray}
\aaop_{\ell m;j,a}&\equiv&
\Big\langle Y_{\ell m}^{\ast}({\bf s}_{j,a})\Big\rangle, 
\label{EQ:vardefseta}
\\
{\rm e}^{-\vert{\bf k}\vert^{2}\xi_{j}^{2}/2}
\pacf_{\ell m;j,a}({\bf k})&\equiv&
\Big\langle
\big(
{\rm e}^{-i{\bf k}\cdot({\bf c}_{j}-{\bfmu}_{j})}-
\langle{\rm e}^{-i{\bf k}\cdot({\bf c}_{j}-{\bfmu}_{j})}\rangle
\big)
\big(
Y_{\ell m}^{\ast}({\bf s}_{j,a})- 
\langle Y_{\ell m}^{\ast}({\bf s}_{j,a})\rangle
\big)
\Big\rangle.
\label{EQ:vardefszeta}
\end{eqnarray}%
\end{mathletters}%
The collection of complex-valued numbers $\{\aaop_{\ell m;j,a}\}$ 
characterizes the orientational localization of orbital $a$ on particle $j$; 
the collection of complex-valued functions 
$\{\pacf_{\ell m;j,a}({\bf k})\}$ 
characterizes the correlations between the thermal fluctuations in the 
position of particle $j$ and the orientation of orbital $a$ attached to it.  
For example, consider $\pacf_{1 0;j,a}({\bf k})$. By expanding the 
exponential to first order (as we shall establish later, typical values of 
$k$ are small near the transition), and recalling, that up to numerical 
factors $Y_{1 0}({\bf s})$ is $s_z$, we see, that
\begin{equation}
{\rm e}^{-\vert{\bf k}\vert^{2}\xi_{j}^{2}/2}\,
\pacf_{1 0;j,a}({\bf k}) = -i \sqrt{{3\over{4 \pi}}} \,\, {\bf k} \cdot
\Big\langle  \big(
({\bf c}_{j}-{\bfmu}_{j}) - 
\langle{\bf c}_{j}-{\bfmu}_{j}\rangle
\big)\big(
({\bf s}_{j,a})_z - \langle({\bf s}_{j,a})_z\rangle
\big)
\Big\rangle + {\cal O}(\hk^2),
\end{equation}
which does indeed measure correlations between the positional and orientational 
thermal fluctuations, in accordance with the discussion
of this component of the order-parameter, given at the end of 
Sec.~\ref{SEC:detector}.

By rewriting Eq.~(\ref{EQ:corr1}) in terms of these diagnostics, and 
making use of the approximation~(\ref{EQ:posapprox}), we arrive at the form
\begin{equation}
\Big\langle{\rm e}^{-i{\bf k}\cdot{\bf c}_{j}}\,
Y_{\ell m}^{\ast}({\bf s}_{j,a})\Big\rangle
\approx
{\rm e}^{-i{\bf k}\cdot{\bfmu}_{j}}\,
 {\rm e}^{-\vert{\bf k}\vert^{2}\xi_{j}^{2}/2}
\Big\{
\aaop_{\ell m;j,a}+\pacf_{\ell m;j,a}({\bf k})
\Big\}.
\label{EQ:corr2}
\end{equation}
By inserting this form, appropriate for particles $j$ that comprise the
localized fraction, into Eq.~(\ref{EQ:NZRop}), and incorporating the 
contribution from the delocalized fraction, we arrive at the form
\begin{eqnarray}
&&
\left[\,
{1\over{N}}\sum_{j=1}^{N}
{1\over{A}}\sum_{a=1}^{A}
\prod_{\alpha=1}^{n}
\Big\langle
{\rm e}^{-i{\bf k}^{\alpha}\cdot{\bf c}_{j}}\,
Y_{\ell^{\alpha} m^{\alpha}}^{\ast}({\bf s}_{j,a})
\Big\rangle
\right]
\nonumber
\\
&&
\qquad\qquad
\approx
(1-q)\prod_{\alpha=1}^{n}\delta_{{\bf k}^{\alpha},{\bf 0}}
+\left[\,
{1\over{N}}{\sum_{j\,{\rm loc.}}}
{1\over{A}}\sum_{a=1}^{A}
\prod_{\alpha=1}^{n}
{\rm e}^{-i{\bf k}^{\alpha}\cdot{\bfmu}_{j}}
{\rm e}^{-\vert{\bf k}^{\alpha}\vert^{2}\xi_{j}^{2}/2}
\big\{\aaop_{\ell^{\alpha}m^{\alpha};j,a}
+\pacf_{\ell^{\alpha}m^{\alpha};j,a}({\bf k}^{\alpha})\big\}
\right].
\\
\nonumber
&&
\qquad\qquad
=
(1-q)\prod_{\alpha=1}^{n}\delta_{{\bf k}^{\alpha},{\bf 0}}
+q\int d^{3}\mu\,
\int d\tau\,d\{\aaop\}\,{\cal D}\{\pacf\}\,
P({\bfmu},\tau,\{\aaop\},\{\pacf\})\,
\\
&&\qquad\qquad\qquad\qquad\times
{\rm e}^{-i{\bfmu}\cdot\sum_{\alpha=1}^{n}{\bf k}^{\alpha}}\,
{\rm e}^{-{1\over{2 \tau}}
	\sum_{\alpha=1}^{n}\vert{\bf k}^{\alpha}\vert^{2}}\,
\prod_{\alpha=1}^{n}
\big\{\aaop_{\ell^{\alpha}m^{\alpha};j,a}
+\pacf_{\ell^{\alpha}m^{\alpha};j,a}({\bf k}^{\alpha})\big\},
\label{EQ:anisoans2}
\end{eqnarray}%
where we have made the definition $\tau \equiv 1/\xi^2$. The integration 
measures $d\{\aaop\}$ and ${\cal D}\{\pacf\}$ respectively denote the 
multiple measure $\prod_{\ell m}d\aaop_{\ell m}$ and the multiple 
functional measure $\prod_{\ell m}{\cal D}\pacf_{\ell m}$. We have also 
introduced the joint probability distribution $P$, central to our 
characterization of the localized particles in amorphous solid state, 
defined via 
\begin{equation}
P({\bfmu},\tau,\{\aaop\},\{\pacf\})\equiv
\left[{1\over{N}}\sum_{j}
{1\over{A}}\sum_{a=1}^{A}
\delta({\bfmu}-{\bfmu}_{j})\,
\delta(\tau-\xi_{j}^{-2})\, 
\prod_{\ell=0}^{\infty}
\prod_{m=-\ell}^{\ell}
\delta(\aaop_{\ell m}-\aaop_{\ell m;j,a})\, 
D(\pacf_{\ell m}-\pacf_{\ell m;j,a})\right], 
\end{equation}
in which the final factor $D(\cdot)$ is a functional delta-function. 

The next step in our construction of a physically motivated form for the
order parameter involves the identification of
specific symmetries that we anticipate the amorphous solid state to
possess, viz., macroscopic translational invariance (MTI) and
macroscopic rotational invariance (MRI).  MTI reflects the notion that
although in the amorphous solid state
translational invariance is spontaneously broken at the
microscopic level (in any particular realization of the disorder), this
invariance is restored at the macroscopic level, in the sense that no
quantity computed by averaging over any macroscopic subvolume of the
system exhibits any dependence on the particular subvolume chosen.  
Similarly, MRI reflects the
notion that although rotational invariance is spontaneously broken at the
microscopic level, it is restored at the macroscopic level in the sense
that no quantity computed by averaging over any macroscopic subvolume of
the system exhibits any orientational preference.

As for MTI, it amounts to the hypotheses: (i)~that the disorder-averaged 
distribution $P$ exhibits no correlation between the mean location of a 
particle and its other statistical characteristics; and 
(ii)~that the distribution is translationally invariant 
(i.e., is independent of ${\bfmu}$).

Although MRI also imposes conditions on the joint probability 
distribution $P$ we do not need to impose these conditions explicitly. 
The reason for that is that
MRI for the component $\fop(\hk;\hz,\hz)$ is assured by the fact that 
$\aaop_{0 0;j,a}$ and $\pacf_{0 0;j,a}$ are constants (in fact, one is zero), 
and the assumption 
that the localization clouds of the particles are spherical and, accordingly,  
described by the single r.m.s.~value of the fluctuation in the particle's position
$3\xi_{j}^2$. MRI for the anisotropic components of $\fop(\hk;\hl,\hm)$ 
is a consequence of the fact that they, as we shall see below, are perturbed 
away from their zero values by MRI-satisfying couplings to the 
$\fop(\hk;\hz,\hz)$ component. Thus MRI is assured by the theory itself, 
and does not need to be explicitly incorporated into the proposed form 
of the order parameter.

By making use of the MTI hypotheses we arrive at the form: 
\begin{equation}
P({\bfmu},\tau,\{\aaop\},\{\pacf\})={P(\tau,\{\aaop\},\{\pacf\})\over{V}},
\end{equation}
which, when inserted into Eq.~(\ref{EQ:anisoans2}), leads to the expression
\begin{eqnarray}
(4\pi)^{n/2}\, 
\fop(\hk;\hl,\hm)\Big\vert_{{\bf k}^{0}={\bf 0},\ell^0=m^0=0}
&=&
(1-q)\prod_{\alpha=1}^{n}\delta_{{\bf k}^{\alpha},{\bf 0}}\,
\delta_{\ell^{\alpha}, 0}\, \delta_{m^{\alpha}, 0}
+q\delta_{\sum_{\alpha=1}^{n}{\bf k}^{\alpha},{\bf 0}}
\int_{0}^{\infty} \!\! d\tau
\int d\{\aaop\}\,{\cal D}\{\pacf\}\,P(\tau,\{\aaop\},\{\pacf\})\,
\nonumber
\\
&&\qquad
\times
{\rm e}^{-{1\over{2\tau}}\sum_{\alpha=0}^{n}
\vert{\bf k}^{\alpha}\vert^{2}}\,
\prod_{\alpha=0}^{n}
\big\{\aaop_{\ell^{\alpha}m^{\alpha};j,a}
+\pacf_{\ell^{\alpha}m^{\alpha};j,a}({\bf k}^{\alpha})\big\},
\label{EQ:anisoans4}
\end{eqnarray}%
where we have introduced the notation 
$\fop(\hk;\hl,\hm)\vert_{{\bf k}^{0}={\bf 0},\ell^0=m^0=0}$ 
to denote the order parameter~(\ref{EQ:NZRop}), and where hats 
indicate $n+1$-fold replicated versions of quantities.  By extending 
the result of this approach to include the dependence on the 
zero-replica variables $({\bf k}^{0},\ell^{0},m^{0})$ in a 
permutation-symmetry--dictated way we arrive at the form
\begin{eqnarray}
(4\pi)^{(n+1)/2}\,
\fop(\hk;\hl,\hm)&=&
(1-q)\prod_{\alpha=0}^{n}\delta_{{\bf k}^{\alpha},{\bf 0}}\,
\delta_{\ell^{\alpha}, 0}\, \delta_{m^{\alpha}, 0}
+q\delta_{\tilde{\bf k},{\bf 0}}
\int_{0}^{\infty}\!\!d\tau
\int d\{\aaop\}\,{\cal D}\{\pacf\}\,p(\tau,\{\aaop\},\{\pacf\})\,
{\rm e}^{-\hk^{2}/2\tau}
\nonumber
\\
&&\qquad\qquad\qquad\qquad
\times
\prod_{\alpha=0}^{n}
\big\{\aaop_{\ell^{\alpha}m^{\alpha};j,a}
+\pacf_{\ell^{\alpha}m^{\alpha};j,a}({\bf k}^{\alpha})\big\}.
\label{EQ:anisoansfin}
\end{eqnarray}%

As we shall see below, for our solution we shall need an assumption for 
the form of the order-parameter component $\fop(\hk;\hz,\hz)$.
To motivate this assumption, 
we set $\hl=\hm=\hz$ in Eq.~(\ref{EQ:anisoansfin}). As is
easy to see from the definitions~(\ref{EQ:vardefseta}) and~(\ref{EQ:vardefszeta}), 
$\aaop_{0 0;j,a}=1/\sqrt{4 \pi}$ and $\pacf_{0 0;j,a}=0$, leaving us with
the form
\begin{equation}
(4\pi)^{(n+1)/2}\,
\fop(\hk;\hz,\hz)=
(1-q)\prod_{\alpha=0}^{n}\delta_{{\bf k}^{\alpha},{\bf 0}}
+q\,\delta_{\tilde{\bf k},{\bf 0}}
\int_{0}^{\infty}\!\!d\tau\,p(\tau)\,
{\rm e}^{-\hk^{2}/2\tau},
\label{EQ:isoansfin}
\end{equation}%
where $p(\tau)$ is a reduced form of the full joint probability distribution 
$P(\tau,\{\aaop\},\{\pacf\})$, and describes only the positional localization of
the particles:
\begin{equation}
p(\tau) \equiv 
\int d\{\aaop\}\,{\cal D}\{\pacf\}\,P(\tau,\{\aaop\},\{\pacf\}).
\label{EQ:reducedpdef}
\end{equation}%
We note that, up to trivial factors of $4 \pi$, the expression 
(\ref{EQ:isoansfin}) is identical to the Ansatz used in 
Refs.~\cite{REF:review,REF:cast} in the context of vulcanized 
macromolecular media.

\section{Disorder-averaging; Replica statistical mechanics}
\label{SEC:RSM}
Having described the relevant \lq\lq kinematics\rlap,\rq\rq\ 
i.e., the degrees of freedom and the constraints that characterize 
the model of randomly covalently bonded particles, we now formulate the 
statistical mechanics of the system, paying particular attention to 
the quenched (i.e., nonequilibrating) nature of the random constraints.  
At this stage we shall be following the method sketched in 
Ref.~\cite{REF:GZeplSG} which itself builds upon the general approach 
to macromolecular networks introduced in Ref.~\cite{REF:demon}.
\subsection{Partition function}
\label{SEC:PartFunc}
The partition function of the system, subject to the constraints 
$\big\{j_{e},j^{\prime}_{e};a_{e},a^{\prime}_{e}\big\}_{e=1}^{M}$,  
which we collectively denote by $\con$,
relative to the partition function of the unconstrained system, 
is given by 
\begin{equation}
\tilde{Z}(\con)
=\Big\langle
\prod_{e=1}^{M}
\delta^{(3)}({\bf c}_{j_{e}}+{1\over{2}}{\bf s}_{j_{e},a_{e}}-
{\bf c}_{j_{e}^{\prime}}-{1\over{2}}{\bf s}_{j_{e}^{\prime},a_{e}^{\prime}})\,
\Delta^{(2)}({\bf s}_{j_{e},a_{e}}+{\bf s}_{j_{e}^{\prime},a_{e}^{\prime}})
\Big\rangle_{N,1}.
\label{EQ:tildeZdef}
\end{equation}
The angle brackets denote equilibrium averaging with respect to a 
Hamiltonian that incorporates interactions between distinct particles,
as well as between the orbitals of a single particle.  The subscript 
indicates that this average is taken over one copy of a system of $N$ 
particles, and anticipates the introduction of replicas that we shall make 
shortly.  The two types of delta-function, $\delta^{(3)}$ and $\Delta^{(2)}$, 
serve to eliminate configurations that fail to satisfy the constraints, 
and are appropriately defined in the following way:
\begin{mathletters}
\begin{eqnarray}
\delta^{(3)}({\bf c}_{1}-{\bf c}_{2})
&=&
\delta^{(3)}({\bf c}_{2}-{\bf c}_{1})
\equiv
\sum_{\bf k}
\left({\exp i{\bf k}\cdot{\bf c}_{1}\over{\sqrt{V}}}\right)
\left({\exp i{\bf k}\cdot{\bf c}_{2}\over{\sqrt{V}}}\right)^{\ast},
\\
\Delta^{(2)}({\bf s}_{1},{\bf s}_{2})
&=&
\Delta^{(2)}({\bf s}_{2},{\bf s}_{1})
\equiv
\sum_{\ell=0}^{\infty}
\sum_{m=-\ell}^{\ell}
Y_{\ell m}       ({\bf s}_{1})\,
Y_{\ell m}^{\ast}({\bf s}_{2}), 
\end{eqnarray}%
\end{mathletters}%
where, (corresponding to the periodic boundary conditions imposed on the system),
the sum over ${\bf k}$ is taken over the Cartesian components
$k_{\nu} = {2 \pi n_{\nu} / V^{1/3}}$, with $n_{\nu}$ being integers (both 
positive and negative), and
the $Y_{\ell m}$ are the usual spherical harmonic functions, the arguments
of which are unit vectors.  As one can readily check by making use of
the orthonormality properties of the spherical harmonic functions, i.e.,
\begin{mathletters}
\begin{eqnarray}
\int_{V}d^{3}c\,
\left({\exp i{\bf k}_{1}\cdot{\bf c}\over{\sqrt{V}}}\right)^{\ast}
\left({\exp i{\bf k}_{2}\cdot{\bf c}\over{\sqrt{V}}}\right)
&=&
\delta_{{\bf k}_{1},{\bf k}_{2}},
\\
\int_{S}d^{2}s\, 
Y_{\ell_{1} m_{1}}^{\ast}({\bf s})\, 
Y_{\ell_{2} m_{2}}       ({\bf s})
&=&
\delta_{\ell_{1},\ell_{2}}\,
\delta_{m_{1},m_{2}},
\end{eqnarray}%
\end{mathletters}%
where the subscript $V$ on the former integral indicates integration 
over the volume of the sample and
the subscript $S$ on the latter integral indicates that the integration
is taken over the two-dimensional surface of a three-dimensional sphere
of unit radius, these delta-functions possess the basic properties:
\begin{mathletters}
\begin{eqnarray}
\int_{V}d^{3}c_{2}\,
\delta^{(3)}({\bf c}_{1}-{\bf c}_{2})\,
\delta^{(3)}({\bf c}_{2}-{\bf c}_{3})
&=&
\delta^{(3)}({\bf c}_{1}-{\bf c}_{3}), 
\label{EQ:dfsym1}
\\
\int_{S}d^{2}s_{2}\, 
\Delta^{(2)}({\bf s}_{1},{\bf s}_{2})\, 
\Delta^{(2)}({\bf s}_{2},{\bf s}_{3})
&=&
\Delta^{(2)}({\bf s}_{1},{\bf s}_{3}). 
\label{EQ:dfsym2}
\end{eqnarray}%
\end{mathletters}

Strictly speaking, the partition function $\tilde{Z}$ is correct only up
to the appropriate Gibbs factorial factor, and would not, as it stands,
give rise to an extensive free energy hence the tilde. As we shall be
focusing on the order-parameter self-consistency equation, in which (as
is well known) the Gibbs factor plays no role it can be safely omitted
here.  For a detailed discussion of this issue, see Sec.~2.4 of
Ref.~\cite{REF:review}.

\subsection{Deam-Edwards distribution}
\label{SEC:DEdist}
At this stage, we introduce a statistical distribution characterizing 
the realization of the random bonds, following the elegant strategy of 
Deam and Edwards~\cite{REF:demon}. 
We take for the probability density that the collection of bonds 
$\con$ is formed the quantity
\begin{equation}
{\cal P}_{M}(\con) 
\propto
{(2\pi V\mu^2/NA^{2})^M\over{M!}}\,
\tilde{Z}(\con),
\label{EQ:bond}
\end{equation}
which is analogous to the probability density chosen by Deam and Edwards 
for the case of vulcanized macromolecular networks~\cite{REF:demon}. 
Instead of working with a fixed number of bonds, we 
allow their number to vary in a quasi-Poissonian way, and control 
the mean number of bonds by the control parameter $\mu^2$. 
For a given value of $M$, the Deam-Edwards distribution is 
proportional to the probability density for finding the 
set of pairs of orbitals 
$\big\{j_{m},j^{\prime}_{m};a_{m},a^{\prime}_{m}\big\}_{m=1}^{M}$ 
to be overlapping.  The factor $\mu^{2M}$ represents the probability 
that a bond is formed between each of these $M$ overlapping pairs. 
Thus, the Deam-Edwards distribution provides a statistical 
characterization of a process of forming permanent bonds in which constraints 
are instantaneously introduced into the liquid state at equilibrium. 
As such, it is an idealization of the random-network-forming process, 
which generally takes place on a time-scale during which at least 
some relaxation of the structure can occur.  To handle the 
complication of relaxation would require the introduction of kinetics into 
the description, rather than purely equilibrium notions.
Said another way, one can view the Deam-Edwards distribution as a 
strategy for freezing in liquid-state correlations, as process that 
is regarded as happening spontaneously in glass-forming systems, 
but here is introduced externally.
The distribution encodes the physically attractive feature that 
the networks it gives appreciable weight to exhibit the macroscopic 
symmetries of the 
liquid state, inasmuch as the bond collections to which it gives 
appreciable weight correspond to likely configurations of the liquid 
state.  With this distribution of bonds some of the correlations 
of the liquid state are {\it quenched in\/}, to a degree controlled 
by the mean number of bonds formed.
\subsection{Replica representation of the disorder-averaged free energy}
\label{SEC:replica}
We now set about constructing the disorder-averaged free energy 
per particle (relative to that of the system prior to random 
covalent bonding) $f$, which is defined via 
\begin{equation}
N\beta f\equiv
\big[\ln \tilde{Z}(\con)\big],  
\end{equation}
where $\beta(\equiv 1/k_{\rm B}T)$ measures the inverse temperature. As 
mentioned in the previous subsection, the Gibbs factor has been omitted, but 
this will have no consequences for the order-parameter self-consistent equation.
By making use of the replica technique (see, e.g., Ref.~\cite{REF:MPVbook}), 
$f$ can be obtained via:
\begin{mathletters}
\begin{eqnarray}
f
&=&
\lim_{n\to 0}f_{n},
\label{EQ:fdef}
\\
-n\beta Nf_{n}
&\equiv&
\big[\tilde{Z}^{n}\big]-1
=\left({\cal Z}_{n+1}-{\cal Z}_{1}\right)/{\cal Z}_{1},
\label{EQ:fndef}
\\
{\cal Z}_{n+1}
&\equiv&
\Bigg\langle
\exp\Bigg(
{2\pi V\mu^{2}
\over{NA^{2}}}
\sum_{j,j'=1}^{N}
\sum_{a,a'=1}^{A}
\prod_{\alpha=0}^{n}
\delta^{(3)}(
 {\bf c}_{j }^{\alpha}+\half{\bf s}_{j ,a }^{\alpha}
-{\bf c}_{j'}^{\alpha}-\half{\bf s}_{j',a'}^{\alpha})\,
\Delta^{(2)}(
 {\bf s}_{j ,a }^{\alpha}, 
-{\bf s}_{j',a'}^{\alpha})
\Bigg)\Bigg\rangle_{N,n+1}.
\label{EQ:zndef}
\end{eqnarray}%
\end{mathletters}%
Here, ${\cal Z}_{n+1}$ is the replicated partition function, arising 
from the averaging of ${\tilde Z}^n$ over the Deam-Edwards--type 
distribution~(\ref{EQ:bond}), and the 
denominator $Z_1$ arises from the normalization 
of the Deam-Edwards distribution (see App.~\ref{APP:DEDaverage} for 
details). Notice the striking occurence of a 
theory involving $n+1$, rather than the usual $n$, replicas, a feature,
arising from the presence in the partition function in the Deam-Edwards 
distribution. (The extra replica ``computes'' the distribution of 
quenched random bonds.)\thinspace The angle 
brackets $\langle\cdots\rangle_{n+1}$ indicate an $n+1$-fold 
replicated normalized average over the positions of the particles and 
the orientations of the orbitals, weighted suitably by a Hamiltonian 
that does not couple the replicas.

As one can see from the exponent in Eq.~(\ref{EQ:zndef}), the replicated 
theory possesses the symmetries of independent translations and 
rotations of the replicas, i.e., 
\begin{mathletters} 
\begin{eqnarray} 
{\bf c}^{\alpha}&\rightarrow&
{\cal R}^{\alpha}\cdot{\bf c}^{\alpha}
+{\bf a}^{\alpha}, 
\\
{\bf s}_{a}^{\alpha}&\rightarrow&
{\cal R}^{\alpha}\cdot{\bf s}_{a}^{\alpha},
\end{eqnarray}%
\end{mathletters}%
where $\{ {\bf a}^{\alpha} \}$ are $n+1$ independent arbitrary 
translation $3$-vectors, and 
$\{{\cal R}^{\alpha}\}$ are $n+1$ independent arbitrary $3$-by-$3$ rotation 
matrices.  As we shall see, the transition to the amorphous solid 
state is marked by the spontaneous breaking of the symmetries of 
the relative translations and rotations of the replicas; the 
common translations and rotations remain as residual symmetries. 
These residual symmetries correspond to the macroscopic translational 
and rotational symmetry of the amorphous solid state discussed in 
Sec.~\ref{SEC:locchardist}. The theory also possesses the symmetry of 
the permutation of the $n+1$ replicas, which remains intact in the 
amorphous solid state.

For the sake of convenience, we introduce the replicated delta-functions,
defined by
\begin{mathletters}
\begin{eqnarray}
\hat{\delta}(\hat{\bf c}_{1}-\hat{\bf c}_{2})
&\equiv&
\prod\nolimits_{\alpha=0}^{n}
\delta^{(3)}({\bf c}_{2}^{\alpha}-{\bf c}_{3}^{\alpha}),
\\
\hat{\Delta}(\hat{\bf s}_{1},\hat{\bf s}_{2})
&\equiv&
\prod\nolimits_{\alpha=0}^{n}
\Delta^{(2)}({\bf s}_{1}^{\alpha},{\bf s}_{2}^{\alpha}), 
\end{eqnarray}%
\end{mathletters}%
where 
$\hat{\bf c}$ denotes $\{{\bf c}^{0},{\bf c}^{1},\ldots,{\bf c}^{n}\}$
and 
$\hat{\bf s}$ denotes $\{{\bf s}^{0},{\bf s}^{1},\ldots,{\bf s}^{n}\}$, 
and also the replicated spherical harmonics $\hy$, defined by
\begin{equation}
\hy_{\hl \hm}(\hs) \equiv
\prod\nolimits_{\alpha=0}^{n} Y_{\ell^{\alpha} m^{\alpha}}({\bf s}^{\alpha}),
\end{equation}%
where $\hl$ and $\hm$ respectively denote $\{\ell^0, \ell^1, \ldots, \ell^n\}$ 
and $\{m^0, m^1, \ldots, m^n\}$.
\section{Mean-field approximation}
\label{SEC:MFAsetup}
\subsection{Self-consistency 
condition for the order parameter}
\label{SEC:MFAderiv}
We now develop a mean-field approximation for the 
replica partition function, Eq.~(\ref{EQ:zndef}).
To do this, we rewrite the partition function as follows:  
\begin{mathletters}
\begin{eqnarray}
{\cal Z}_{n+1}
&=&
\left\langle\exp
\bigg(
2\pi NV\mu^{2}\int_{V}d\hat{c}\,\int_{S}d\hat{s}\,
{1\over{NA}}\sum_{j=1}^{N}\sum_{a=1}^{A}
\hat{\delta}
	\big((\hat{\bf c}_{j }+\half\hat{\bf s}_{j ,a })-\hat{\bf c}\big)\,
\hat{\Delta}
	\big(\hat{\bf s}_{j,a},\hat{\bf s}\big)
\right.
\nonumber\\
&&\qquad\qquad\qquad\qquad
\times
\left.
{1\over{NA}}\sum_{j'=1}^{N}\sum_{a'=1}^{A}
\hat{\delta}
	\big(\hat{\bf c}-(\hat{\bf c}_{j'}+\half\hat{\bf s}_{j',a'})\big)\,
\hat{\Delta}
	\big(\hat{\bf s},-\hat{\bf s}_{j',a'}\big)
\bigg)\right\rangle_{N,n+1}
\\ 
&=&
\left\langle\exp
\bigg(
2\pi NV\mu^{2}\int_{V}d\hat{c}\,\int_{S}d\hat{s}\,
{1\over{NA}}\sum_{j=1}^{N}\sum_{a=1}^{A}
\hat{\delta}
	\big(\hat{\bf c}-(\hat{\bf c}_{j }+\half\hat{\bf s}_{j ,a })\big)\,
\hat{\Delta}
	\big(\hat{\bf s},\hat{\bf s}_{j,a}\big)
\right.
\nonumber\\
&&\qquad\qquad\qquad\qquad
\times
\left.
{1\over{NA}}\sum_{j'=1}^{N}\sum_{a'=1}^{A}
\hat{\delta}
	\big(\hat{\bf c}-(\hat{\bf c}_{j'}+\half\hat{\bf s}_{j',a'})\big)\,
\hat{\Delta}
	\big(-\hat{\bf s},\hat{\bf s}_{j',a'}\big)
\bigg)\right\rangle_{N,n+1}, 
\label{EQ:lastNpcle}
\end{eqnarray}%
\end{mathletters}%
where $\int_{V}d\hat{c}$ denotes 
$\prod_{\alpha=0}^{n}\int_{V}d^{3}{\bf c}^{\alpha}$, 
and $\int_{S}d\hat{s}$ denotes 
$\prod_{\alpha=0}^{n}\int_{S}d^{2}{\bf s}^{\alpha}$, 
and where, to obtain the last form, we have used the symmetry property 
of the delta-functions, Eqs.~(\ref{EQ:dfsym1},\ref{EQ:dfsym2}).

Next, we introduce the (real-space version of the) amorphous solid 
order parameter,
\begin{equation}
\Omega(\hat{\bf c};\hat{\bf s}) \equiv
\left\langle
{1\over{NA}}\sum_{j=1}^{N}\sum_{a=1}^{A}
\hat{\delta}
	\big(\hat{\bf c}-(\hat{\bf c}_{j }+\half\hat{\bf s}_{j,a})\big)\,
\hat{\Delta}
	\big(\hat{\bf s},\hat{\bf s}_{j,a}\big) 
\right\rangle. 
\end{equation}Then, 
upon setting 
\begin{equation}
{1\over{NA}}\sum_{j=1}^{N}\sum_{a=1}^{A}
\hat{\delta}
	\big(\hat{\bf c}-(\hat{\bf c}_{j }+\half\hat{\bf s}_{j,a})\big)\,
\hat{\Delta}
	\big(\hat{\bf s},\hat{\bf s}_{j,a}\big)
=     \Omega(\hat{\bf c};\hat{\bf s})+
\delta\Omega(\hat{\bf c};\hat{\bf s}),
\end{equation}
i.e., the order parameter $\fop(\hat{\bf c};\hs)$ plus the fluctuation 
$\delta\Omega(\hat{\bf c};\hat{\bf s})$,
expanding the exponent in powers of $\delta\Omega(\hat{\bf c};\hat{\bf s})$, 
and omitting terms quadratic in $\delta\Omega(\hat{\bf c};\hat{\bf s})$, 
we obtain
\begin{eqnarray}
{\cal Z}_{n+1}
&\approx&
\exp\left(
-2\pi NV\mu^{2}
\int_{V}\!d\hat{c}\,\int_{S}\!d\hat{s}\,
\Omega(\hat{\bf c}; \hat{\bf s})\,
\Omega(\hat{\bf c};-\hat{\bf s})
\nonumber
\right.
\\
&&\left.\quad
+N\ln\Bigg\langle\exp\Bigg(
4\pi V\mu^{2}\int_{V}\!d\hat{c}\,\int_{S}\!d\hat{s}\,\,
\Omega(\hat{\bf c};-\hat{\bf s})
{1\over{A}}\sum_{a=1}^{A}
\hat{\delta}
	\big(\hat{\bf c}-(\hat{\bf c}_{1}+\half\hat{\bf s}_{1,a})\big)\,
\hat{\Delta}
	\big(\hat{\bf s},\hat{\bf s}_{1,a}\big)
\Bigg)\Bigg\rangle_{1,n+1}\right),  
\label{EQ:partfun}
\end{eqnarray}%
where the resulting expectation value involves only the position and
orbital-orientations of a single particle.  

The reader will have observed that the mean-field approximation strategy
has yielded a one-particle problem, Eq.~(\ref{EQ:partfun}), as desired.
However, there is a subtlety associated with the manner in which the
various interactions present in Eq.~(\ref{EQ:lastNpcle}) are treated,
which we now address.  The angle brackets in Eq.~(\ref{EQ:lastNpcle})
denote averaging over $n+1$ (coupled) replicas of the $N$ (coupled)
particle system.  The intra-replica coupling originates in the
interactions between particles present in the liquid state; on the other
hand, the inter-replica coupling originates in the random constraints.
As discussed in detail in Sec.~5.1 of Ref.~\cite{REF:review}, it is
useful to transfer the so-called one-replica sector contribution to the
inter-replica coupling to the intra-replica coupling (which is thereby
renormalized). (The intra- and inter-replica couplings both contain
trivial contributions in the zero-replica sector, as does the order
parameter; we ignore these contributions.)\thinspace\  The subtlety is
that the structure of the theory in the one-replica sector is quite
different from that in the higher-replica sectors: whereas the
constraints tend to destabilize {\it all\/} sectors, this tendency is
counteracted in only the one-replica sector by the original
intra-replica interactions.  Consequently, at the amorphous
solidification transition the one-replica sector component of the order
parameter remains zero, whilst the higher-replica sector components
become nonzero. Indeed, the competition between these two processes can
be regarded as a form of frustration, which resolves itself by the
formation of a state possessing MTI and MRI (see
Sec.~\ref{SEC:locchardist}).

On a technical level, this discussion amounts to the following dictum:
in all subsequent equations, e.g.,
Eqs.~(\ref{EQ:partfun},\ref{EQ:SCEReal},\ref{EQ:SCE}), the component of
the order parameter lying in the one-replica sector is to be set to
zero.  Accordingly, the self-consistent equations that follow pertain to
all sectors {\it except\/} the zero and one replica sectors. This notion
is straightforward when the order parameter is expressed in the (plane
and spherical) harmonic representation [as it is, e.g., in
Eq.~(\ref{EQ:SCE})]. In this representation, setting the one-replica
sector contribution to zero refers to setting to zero  the contribution
in which nonzero entries in $\{\hk;\hl,\hm\}$ appear in precisely one
replica.  (By the zero-replica sector we mean the sector with
$\hk=\hl=\hm=\hz$.)

We now return to the task of obtaining a self-consistent equation for
the order parameter.  By making the partition
function~(\ref{EQ:partfun}) stationary with respect to 
$\Omega(\hat{\bf c};\hat{\bf s})$ we arrive at
self-consistent equation (\sce ) for the order-parameter:
\begin{eqnarray}
\Omega(\hat{\bf c}_0;\hs_0)
&=&
{\displaystyle
{\displaystyle
\left\langle
{1\over{A}}\sum_{a=1}^{A}
\hat{\delta}
	\big(\hat{\bf c}_0-(\hat{\bf c}_{1}+\half\hs_{1,a})\big)\,
\hat{\Delta}
	\big(\hat{\bf s}_0,\hs_{1,a}\big)
\right.
\hfill\atop{\displaystyle
\quad\qquad
\left.
\times\exp\left(
4\pi V\mu^{2}\int_{V}\!d\hat{c}\,\int_{S}\!d\hat{s}\,\,
\Omega(\hat{\bf c};-\hat{\bf s})
{1\over{A}}\sum_{a=1}^{A}
\hat{\delta}
	\big(\hat{\bf c}-(\hat{\bf c}_{1}+\half\hs_{1,a})\big)\,
\hat{\Delta}
	\big(\hat{\bf s},\hs_{1,a}\big)
\right)\right\rangle_{1,n+1}
}}
\over{\displaystyle
\left\langle\exp\left(
4\pi V\mu^{2}\int_{V}\!d\hat{c}\,\int_{S}\!d\hat{s}\,\,
\Omega(\hat{\bf c};-\hs)
{1\over{A}}\sum_{a=1}^{A}
\hat{\delta}
	\big(\hat{\bf c}-(\hat{\bf c}_{1}+\half\hs_{1,a})\big)\,
\hat{\Delta}
	\big(\hs,\hs_{1,a}\big)
\right)\right\rangle_{1,n+1}
}}. 
\label{EQ:SCEReal}
\end{eqnarray}%
The presence of $\hat{\Delta}$-function factors provides
the option of replacing the dynamical variable $\hat{\bf s}_{1,a}$ 
in the argument of the $\delta$-functions by the parametric variable 
$\hat{\bf s}$, which replacement we sometimes make.

The transformation of the order parameter to a representation in terms
of the plane wave and spherical harmonic coordinates, and the inverse
transformation, are effected as follows:
\begin{mathletters}
\begin{eqnarray}
\Omega(\hk;\hat{\ell},\hat{m})
&=&
\int_{S}d\hat{s}\,\,
\hy_{\hl \hm}^{\ast}(\hs)\,
\exp(\half i \hk\cdot\hs)
\int_{V}d\hat{c}\,
\exp(-i\hk\cdot \hat{\bf c})\,
\Omega(\hat{\bf c};\hs),
\\
\Omega(\hat{\bf c};\hs)
&=&
{1\over{V^{n+1}}}
\sum\nolimits_{\hat{\bf k}}
\exp(i\hk\cdot\hat{\bf c})
\exp(-\half i \hk\cdot\hs)
\sum\nolimits_{\hat{\ell}\hat{m}}
\hy_{\hl \hm} (\hs)\,
\Omega(\hat{\bf k};\hat{\ell},\hat{m}).
\end{eqnarray}%
\end{mathletters}%
This choice of transformation has the effect of 
keeping the physical interpretation of the order parameter 
$\fop(\hk;\hl,\hm)$ free of any complicating factors of 
$\exp\big({\half i \hat{\bf k}\cdot\hat{\bf s}}\big)$, 
leaving the full factor 
$\exp\big({i\hat{\bf k}\cdot\hat{\bf s}}\big)$ 
in the fluctuating variable to which the order parameter couples.  
Via this transformation, one arrives at a transformed self-consistent
equation for the order parameter:
\begin{eqnarray}
\fop(\hk_0;\hl_0,\hm_0)
&=&
{
{\left\langle
{1\over{A}}\sum_{a=1}^{A}
{\rm e}^{-i\hat{\bf k}_0\cdot\hat{\bf c}}\,
\hy^{\ast}_{\hl_0 \hm_0} (\hs_a)
\exp\left(
{4\pi\mu^{2}\over {V^{n}}}
\sum_{\hk,\hl,\hm}
\fop(\hk;\hl,\hm)\,
{\rm e}^{i\hat{\bf k}\cdot\hat{\bf c}}\,
{1\over{A}}\sum_{a=1}^{A}
{\rm e}^{i\hat{\bf k}\cdot\hat{\bf s}_{a}}\,
\hy_{\hl \hm} (-\hs_a)
\right)
\right\rangle_{1,n+1}
}
\over{
\left\langle
\exp
\left(
{4\pi\mu^{2}\over {V^{n}}}
\sum_{\hk,\hl,\hm}
\fop(\hk;\hl,\hm)\,
{\rm e}^{i\hat{\bf k}\cdot\hat{\bf c}}\,
{1\over{A}}\sum_{a=1}^{A}
{\rm e}^{i\hat{\bf k}\cdot\hat{\bf s}_{a}}\,
\hy_{\hl \hm} (-\hs_a)
\right)
\right\rangle_{1,n+1} 
}}.
\label{EQ:SCE}
\end{eqnarray}%
We shall often refer to the triple $(\hk_0;\hl_0,\hm_0)$ as 
the ``external variables\rlap.''
\subsection{Instability of the fluid state}
\label{SEC:MFAls}
We now demonstrate that upon increasing the density of formed bonds 
the fluid state is rendered linearly unstable.  To do this we follow
Ref.~\cite{REF:alloys}, and expand the replica free energy, 
$n \beta N f_n$ in Eq.~(\ref{EQ:fndef}), to second 
order in the order parameter $\fop$, thus obtaining
\begin{eqnarray}
n \beta N f_n (\fop) &\approx& 
2 \pi \mu^2 N V \sum_{\hk} \sum_{\hl_1, \hm_1} \sum_{\hl_2,\hm_2}
\fop(\hk;\hl_1,\hm_1)\, 
\fop(-\hk;\hl_2,\hm_2)\, 
M_{\hl_1 \hm_1,\hl_2 \hm_2}(\hk)
\nonumber
\\
&&\qquad 
- \left(
{4 \pi \mu^2 \over{A}}
\right)^2 
{ N V \over {2 V^n}}
\sum_{\hk,\hl,\hm} 
\left\vert \fop(\hk;\hl,\hm)\right\vert^2 
(-1)^{\tilde{\ell}}
\sum_{a_1,a_2} \prod_{\alpha=0}^{n} 
\left(
{\delta_{a_1,a_2} + \left(1-\delta_{a_1,a_2}\right)
\co{\ell^{\alpha}}\over {4 \pi}}
\right),
\label{EQ:stability}
\end{eqnarray}%
where, as discussed in Sec.~\ref{SEC:elements}, 
the coefficients $\co{\ell}$ represent the effects of the interactions
between the orbitals of a single particle (and are all less than $1$ for 
$\ell \ge 1$), and the kernel $M$ is given by
\begin{equation}
M_{\hl_1 \hm_1,\hl_2 \hm_2}(\hk) = 
\int_{S} d\hs \, \, \hy_{\hl_1 \hm_1} (\hs)\,
 \hy_{\hl_2 \hm_2} (-\hs) \, \, 
\exp(i \hk \cdot \hs).
\end{equation}
We remind the reader that this linear stability analysis applies only to
the higher replica sectors of the order parameter, as per the discussion
in Sec.~\ref{SEC:MFAderiv}.  The corresponding analysis applied to the
one-replica sector reveals the fact, anticipated in
Sec.~\ref{SEC:MFAderiv}  that the one-relica sector remains stable at
the transition, owing to the stabilizing effect of the inter-particle
interactions.

We expect the liquid state to become unstable first for long wavelengths,
corresponding to $\hk \to {\bfhz}$. In this limit,
$M_{\hl_1 \hm_1,\hl_2 \hm_2}(\hk) \to \delta_{\hl_1,\hl_2} 
\delta_{\hm_1,\hm_2} (-1)^{\tilde{\ell}+\tilde{m}}$.
By examining the coefficients of the quadratic terms in 
Eq.~(\ref{EQ:stability}), and specifically their signs, we see that 
for $\mu^2<\mu^2_c\equiv1$ the coefficients are positive 
for all components of the order parameter [i.e., for all values of 
$(\hk,\hl,\hm)$] and, therefore, that the free energy has a local
minimum at $\fop=0$.  Thus, for $\mu^2 \le \mu^2_{\rm c}$ the fluid state is (at least 
locally) thermodynamically stable.  On the other hand, for 
$\mu^2>\mu_{\rm c}^2$, certain coefficients become negative, starting with 
the longest length-scale (and isotropic, corresponding to $\hl = \hz$) 
modes.  This sign-change  
indicates the loss of the (linear) stability of the fluid, and the 
concomitant acquisition of a nonzero value of the order parameter. 
As usual, the linear instability of one state does not sharply 
specify the nature of the stable state that replaces it, 
although the directions of instability do provide hints. 
In the present setting, the residual stability of the 
one-replica sector suggests that the primary characteristic of the 
new state is macroscopic translational and rotational invariance, 
which is the mechanism by which the induction of energetically-costly 
order in the one-replica sector is avoided.  Thus, it is reasonable 
to anticipate that the state that replaces the fluid state upon the 
formation of a sufficiently large density of bonds is the amorphous 
solid state.  Furthermore, as the coefficients $\co{\ell}$ are smaller
than unity for 
$\ell \ge 1$, and become progressively smaller with increasing $\ell$, 
we may conclude that all anisotropic sectors remain stable, 
at least for bond densities in the vicinity of the amorphous 
solidification transition. Thus, it is reasonable to anticipate that 
anisotropic ordering (i.e., orientational localization) will 
arise only as a {\it response} to positional ordering, via nonlinear 
coupling between isotropic and anisotropic order-parameter components. 
Thus, as will be borne out below, we should anticipate that the form and
extent of the anisotropic ordering will be computable algorithmically, 
as a perturbative correction to the nonperturbative result for the 
ordering in the isotropic sector \cite{REF:stability}.
\section{Solution of the order-parameter self-consistent equation}
\label{SEC:SCEsol}
\subsection{General strategy}
\label{SEC:strategy}
Having obtained the self-consistency equation (\sce ) for the order parameter, 
Eq.~(\ref{EQ:SCE}), we now turn to the issue of solving it. We shall begin by
extracting from the full \sce\ a transcendental self-consistency equation 
for the fraction of localized particles $q$ by considering the \sce\ at external
variables $\hl_0=\hm_0=\hz$ and taking the limit $\hk_0 \to {\bfhz}$. 
Solving this equation in the vicinity of the amorphous solidification 
transition (i.e., for small excess crosslink densities) we will find
that $q$ tends to $0$ near the transition, allowing us to expand 
the \sce\ for the order-parameter in powers of $q$ and truncate the expansion, 
retaining terms of order $q^2$. 
\begin{figure}[hbt]
 \epsfxsize=3.0in
 \centerline{\epsfbox{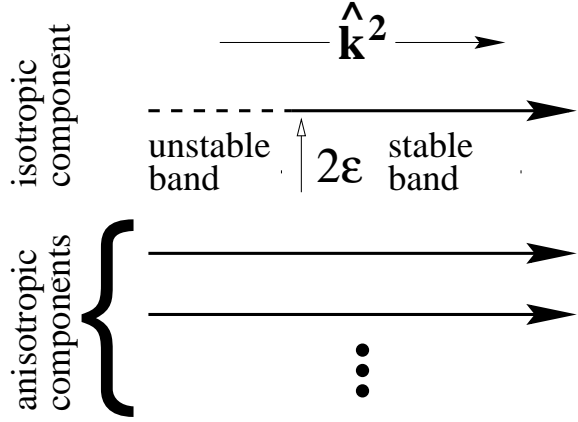}}
\vskip0.5truecm
\caption{
Component- and band-structure of the order parameter (see text for 
explanation).}
\label{FIG:rail_roa}
\end{figure}%
We will then solve the \sce\ for individual order parameter components,
starting with the only component with an unstable band, the isotropic 
component $\fop(\hk;\hz,\hz)$ (the unstable band being those long 
wavelength modes for which $\hk^2 < 2 \cp$), 
which we shall obtain by utilizing the form of the solution,
due to refs.~\cite{REF:cast,REF:review}, and
discussed in Sec.~\ref{SEC:locchardist}. The remaining (anisotropic) 
components are 
linearly stable and thus we can solve for their leading-order values 
perturbatively, by 
considering their couplings to $\fop(\hk;\hz,\hz)$. 
We shall obtain these anisotropic 
components by first solving for the leading-order contribution to the
two lowest angular-momentum components, 
and then obtaining the solution in the general case by induction.
The structure of the order parameter is illustrated in 
Fig.~\ref{FIG:rail_roa}, showing the different components and bands.

Throughout the entire calculation we shall only be concerned with finding the
leading-order contributions to the components of $\fop$. This will generally imply 
(i)~truncating the expansion in powers of $q$ (typically at second order), 
(ii)~ignoring the coupling of the components of the order parameter to
higher angular-momentum components, and (iii)~truncating expansions in powers
of $\hk$ (typically at linear order), as we expect typical values of $\hk^2$ to 
be of order $\cp$. Many technical details of the calculations have been 
relegated to the appendices.
\subsection{Fraction of positionally localized particles}
\label{SEC:gelfraction}
The first step in our solution of the order-parameter \sce\ is to determine the
fraction of localized particles $q$. 
Following the discussion in Sec.~\ref{SEC:LimitFraction}, we first
separate the delocalized and localized fractions in 
the full order parameter $\fop$ by writing
\begin{equation}
\fop(\hk;\hl,\hm)\equiv
(1-q)\,\prod_{\alpha=0}^{n}
\left(
{\delta_{{\bf k}^{\alpha},{\bf 0}}\,
\delta_{\ell^{\alpha}, 0}\, \delta_{m^{\alpha}, 0} \over {\sqrt{4 \pi}}}
\right)
+q \pop(\hk;\hl,\hm),
\label{EQ:popdef}
\end{equation}
where $q\,\pop(\hk;\hl,\hm)$ is the part of the order parameter describing the
localized particles and is analytic at the origin, with 
$\pop(\bfhz;\hz,\hz)= 1/\sqrt{4 \pi}$
The delocalized contribution cancels from the numerator and
denominator of the \sce , so we may rewrite 
Eq.~(\ref{EQ:SCE}), replacing $\fop(\hk;\hl,\hm)$ with 
$q \pop(\hk;\hl,\hm)$ in both the numerator and the denominator, obtaining
\begin{eqnarray}
\fop(\hk_0;\hl_0,\hm_0)
&=&
{
{\left\langle
{1\over{A}}\sum_{a=1}^{A}
{\rm e}^{-i\hat{\bf k}_0\cdot\hat{\bf c}}\,
\hy^{\ast}_{\hl_0 \hm_0} (\hs_a)
\exp\left(
{4\pi\mu^{2}\over {V^{n}}}
\sum_{\hk,\hl,\hm}
q\,\pop(\hk;\hl,\hm)\,
{\rm e}^{i\hat{\bf k}\cdot\hat{\bf c}}\,
{1\over{A}}\sum_{a=1}^{A}
{\rm e}^{i\hat{\bf k}\cdot\hat{\bf s}_{a}}\,
\hy_{\hl \hm} (-\hs_a)
\right)
\right\rangle_{1,n+1}
}
\over{
\left\langle
\exp
\left(
{4\pi\mu^{2}\over {V^{n}}}
\sum_{\hk,\hl,\hm}
q\,\pop(\hk;\hl,\hm)\,
{\rm e}^{i\hat{\bf k}\cdot\hat{\bf c}}\,
{1\over{A}}\sum_{a=1}^{A}
{\rm e}^{i\hat{\bf k}\cdot\hat{\bf s}_{a}}\,
\hy_{\hl \hm} (-\hs_a)
\right)
\right\rangle_{1,n+1} 
}}.
\label{EQ:SCEc}
\end{eqnarray}%

Following the ideas of 
Refs.~\cite{REF:review,REF:cast}, to obtain a \sce\ for $q$ we 
consider the \sce\ (\ref{EQ:SCEc}) for the order-parameter component
$\fop(\hk_0;\hz,\hz)$ in the limit $\hk_0 \to {\bfhz}$. We start with the
\sce\ for $\fop(\hk_0;\hz,\hz)$:
\begin{equation}
\fop(\hk_0;\hz,\hz) = 
{\rm e}^{-\mu^2 q}\,
\left\langle
{\rm e}^{-i\hat{\bf k}_0\cdot\hat{\bf c}}\,
{1\over{\sqrt{4 \pi}}}^{n+1}
\exp\bigg(
{4\pi\mu^{2}\over {V^{n}}}
\sum_{\hk,\hl,\hm}
q\,\pop(\hk;\hl,\hm)\,
{\rm e}^{i\hat{\bf k}\cdot\hat{\bf c}}\,
{1\over{A}}\sum_{a=1}^{A}
{\rm e}^{i\hat{\bf k}\cdot\hat{\bf s}_{a}}\,
\hy_{\hl \hm} (-\hs_a)
\bigg)
\right\rangle_{1,n+1}.
\end{equation}
Note that the denominator of the right-hand side of Eq.~(\ref{EQ:SCEc}) has
been replaced by $\exp (\mu^2 q)$ (see App.~\ref{APP:denominator} for
details). We now consider the limit $\hk \to {\bfhz}$ via a sequence for which
$\tilde{\bf k} = {\bf 0}$. The left-hand side,
as can be easily seen from Eq.~(\ref{EQ:popdef}), becomes 
$q / \sqrt{4 \pi}$. To evaluate the right-hand
side we follow the procedure used in App.~\ref{APP:denominator} to obtain
the denominator of the right-hand side of the \sce .

We expand the exponential in a power series in $q$ and consider the $r$-th order term,
and pass to the replica limit $n \to 0$:
\begin{eqnarray}
&&
\left\langle {1\over{r !}}
\left(4\pi\mu^{2}\right)^{r} q^r
\sum_{{\bf k}_1 \ell_1 m_1}
\ldots
\sum_{{\bf k}_r \ell_r m_r}
\pop({\bf k}_1; \ell_1, m_1)\,
\ldots
\pop({\bf k}_r; \ell_r, m_r)\,
{\rm e}^{i {\bf k}_0\cdot {\bf c}}\,
{\rm e}^{i {\bf k}_1\cdot {\bf c}}\,
\ldots
{\rm e}^{i {\bf k}_r\cdot {\bf c}}\,
\right.
\nonumber
\\
&&\qquad\qquad\qquad\times\left.
{1\over{A^r}} \sum_{a_1, \ldots, a_r}
{\rm e}^{i {\bf k}_1\cdot {\bf s}_{{a}_1}}
\ldots
{\rm e}^{i {\bf k}_r\cdot {\bf s}_{{a}_r}}
Y_{\ell_1 m_1 }(-{\bf s}_{a_1})
\ldots
Y_{\ell_r m_r}(-{\bf s}_{a_r})
\right\rangle_{1,1} .
\label{EQ:num3}
\end{eqnarray}%
By MTI of $\pop(\hk; \hl, \hm)$ we know that $\tilde{\bf k}_i = {\bf 0}$, 
which means that ${\bf k}_i = {\bf 0}$.  
Also, using MRI of $\pop(\hk; \hl, \hm_)$
we see that if $\hk = \hz$ then (\ref{EQ:denom3}) can only be nonzero if
$\hl=\hz$ also. 
Knowing that $\pop(\hz; \hz, \hz)=1/\sqrt{4 \pi}$, for the 
$r$-th order contribution  (with $r\ge 1$) we get 
${1\over{r !}} {(\mu^2)}^r q^r$. 
For $r=0$, however, we get $0$, becausethe limit of 
$\langle \exp ( - i {\bf k}_0 \cdot {\bf c}) \rangle$ 
as ${\bf k}_0 \to {\bf 0}$ is zero.  
Resumming the power series we finally obtain
\begin{equation}
{\displaystyle
\left\langle
{\rm e}^{i\hat{\bf k}_0\cdot\hat{\bf c}}\,
\exp\bigg(
{4\pi\mu^{2} \over {V^{n}}}
\sum_{\hat{\bf k}\hat{\ell}\hat{m}}
q\,\pop(\hk;\hl,\hm)\,
{\rm e}^{i\hat{\bf k}\cdot\hat{\bf c}}\,
{1\over{A}}\sum_{a=1}^{A}
{\rm e}^{i\hat{\bf k}\cdot\hat{\bf s}_{a}}
\hy_{\hl \hm}(\hs_{a})
\bigg)\right\rangle_{1,n+1} }
= \sum_{r>1} {1\over{r !}} {(\mu^2)}^r q^r
= \exp(\mu^2 q) - 1. 
\label{EQ:numf}
\end{equation}
We thus find the following self-consistency equation for 
the fraction of localized atoms:
\begin{equation}
1-q=\exp(-\mu^2 q).
\label{EQ:fracsce}
\end{equation}%
This self-consistency condition is precisely that obtained in the 
case of vulcanized macromolecules~\cite{REF:cast,REF:review}, 
and earlier, in the context of random graph theory, by 
Erd\H os and R\'enyi (see Ref.~\cite{REF:review}).

We will find it convenient to exchanging the control parameter 
$\mu^2$ for $\cp$, defined via $\mu^2 \equiv 1 + \cp/3$, 
so that $\cp$ vanishes as the transition is approached.
The self-consistent equation for $q$ is transcendental, but it is easy to
analyze it graphically. Then, for $\cp < 0$ we find that there is only one
solution $q=0$, corresponding to the liquid state of the system (no localized
particles), and for $\cp > 0$ we find that for $\cp$ small $q$ is small also,
indicating that the fraction 
$q$ of particles comprising the amorphous solid state
tends to zero in the vicinity of the amorphous solidification transition. 
We can thus expand the exponential on 
the right-hand side of Eq.~(\ref{EQ:fracsce})
obtaining the fraction $q$ to first order in $\cp$:
\begin{eqnarray}
q \approx {2\over{3}} \cp.
\label{EQ:qeval}
\end{eqnarray}%
\subsection{Perturbation expansion for the self-consistency equation}
\label{SEC:perturb}
Having found the fraction $q$ of localized particles to be small in the vicinity
of the transition, we may expand the self-consistency equation for the order
parameter, Eq.~(\ref{EQ:SCEc}), in powers of $q$, to second order, obtaining:
\begin{eqnarray}
q \, \pop(\hk_{0};\hl_{0},\hm_{0}) &+&
(1-q)\,\prod_{\alpha=0}^{n}
\left(
{\delta_{{\bf k}^{\alpha},{\bf 0}} \,
\delta_{\ell^{\alpha}, 0}\, \delta_{m^{\alpha}, 0} \over {\sqrt{4 \pi}}}
\right) 
\nonumber
\\
\qquad
\approx &\phantom{+}&
{\rm e}^{- \mu^2 q} \,
\left\langle
{1\over{A}}\sum_{a_{0}=1}^{A}
{\rm e}^{-i\hk_{0}\cdot\hat{\bf c}}\,
\hy^{\ast}_{\hl_0 \hm_0}(\hs_{a_0})
\right\rangle_{1,n+1}
\nonumber
\\
&+& {\rm e}^{- \mu^2 q} \,
\left\langle
{1\over{A}}\sum_{a_{0}=1}^{A}
{\rm e}^{-i\hk_{0}\cdot\hat{\bf c}}\,
\hy^{\ast}_{\hl_0 \hm_0}(\hs_{a_0})
{4\pi\mu^{2}\over{V^{n}}}
\sum_{\hk_1,\hl_{1},\hm_{1}}\!\!
q \, \pop(\hk_{1};\hl_{1},\hm_{1})\,
{\rm e}^{i\hk_{1}\cdot\hat{\bf c}}
{1\over{A}}\sum_{a_{1}=1}^{A}
{\rm e}^{i\hk_{1}\cdot\hat{\bf s}_{a_{1}}}\,
\hy_{\hl \hm}(-\hs_{a_1})
\right\rangle_{1,n+1}
\nonumber
\\
&+& {{\rm e}^{- \mu^2 q}\over{2}}\,
\left\langle
{1\over{A}}\sum_{a_{0}=1}^{A}
{\rm e}^{-i\hk_{0}\cdot\hat{\bf c}}\,
\hy^{\ast}_{\hl_0 \hm_0}(\hs_{a_0})
{4\pi\mu^{2}\over{V^{n}}}
\sum_{\hk_1,\hl_{1},\hm_{1}}\!\!
q\,\pop(\hk_{1};\hl_{1},\hm_{1})\,\,
{\rm e}^{i\hk_{1}\cdot\hat{\bf c}}\,
{1\over{A}}\sum_{a_{1}=1}^{A}
{\rm e}^{i\hk_{1}\cdot\hat{\bf s}_{a_{1}}}\,
\hy_{\hl \hm}(-\hs_{a_1})
\right.
\nonumber
\\
&&\qquad\qquad\times
\left.
{4\pi\mu^{2}\over{V^{n}}}
\sum_{\hk_2,\hl_{2},\hm_{2}}
q\,\pop(\hk_{2};\hl_{2},\hm_{2})\,
{\rm e}^{i\hk_{2}\cdot\hat{\bf c}}\,
{1\over{A}}\sum_{a_{2}=1}^{A}
{\rm e}^{i\hk_{2}\cdot\hat{\bf s}_{a_{2}}}\,
\hy_{\hl \hm}(-\hs_{a_2})
\right\rangle_{1,n+1}
\end{eqnarray}%
Cancelling the liquid contribution to $\fop$ on the left-hand side with 
the 0-th order
contribution on the right-hand side,
rearranging terms and replacing $\exp(-\mu^2 q)$ with $1-q$ we arrive 
at the form of the SCE that we shall focus on:
\begin{eqnarray}
q\,\pop(\hk_0;\hl_0,\hm_0)&\approx&
(1-q) \,
{4\pi\mu^{2}\over{V^{n}}}\,q
\sum_{\hk_1,\hl_{1},\hm_{1}}
\!\!\pop(\hk_{1};\hl_{1},\hm_{1})\,
{1\over{A^{2}}}
\sum_{a_{0},a_{1}=1}^{A}
\left\langle
{\rm e}^{-i\hk_{0}\cdot\hat{\bf c}}\,
\hy^{\ast}_{\hl_0 \hm_0}(\hs_{a_0})\,
{\rm e}^{i\hk_{1}\cdot\hat{\bf c}}\,
{\rm e}^{i\hk_{1}\cdot\hat{\bf s}_{a_{1}}}\,
\hy_{\hl \hm}(-\hs_{a_1})
\right\rangle_{1,n+1}
\nonumber
\\
&&
\qquad + {1-q \over {2}}
\left({4\pi\mu^{2}\over{V^{n}}}\right)^{2}\,q^{2}
\sum_{\hk_1,\hl_{1},\hm_{1}}
\sum_{\hk_2,\hl_{2},\hm_{2}}
\!\!\pop(\hk_{1};\hl_{1},\hm_{1})\,
\pop(\hk_{2};\hl_{2},\hm_{2})
\nonumber
\\
&&\times
{1\over{A^{3}}}\sum_{a_{0},a_{1},a_{2}=1}^{A}
\left\langle
{\rm e}^{-i\hk_{0}\cdot\hat{\bf c}}\,
\hy^{\ast}_{\hl_0 \hm_0}(\hs_{a_0})\,
{\rm e}^{i\hk_{1}\cdot\hat{\bf c}}\,
{\rm e}^{i\hk_{1}\cdot\hat{\bf s}_{a_{1}}}\,
\hy_{\hl \hm}(-\hs_{a_1})\,
{\rm e}^{i\hk_{2}\cdot\hat{\bf c}}\,
{\rm e}^{i\hk_{2}\cdot\hat{\bf s}_{a_{2}}}\,
\hy_{\hl \hm}(-\hs_{a_2})
\right\rangle_{1,n+1}.
\label{EQ:SCEexp}
\end{eqnarray}%
\subsection{Self-consistency equation: Isotropic sector}
\label{SEC:isosector}
Having obtained the localized fraction $q$ and verified the consistency of
the expansion of the \sce, we now turn to the issue of solving the \sce\ 
for individual components of the order-parameter.

As discussed in Sec.~\ref{SEC:MFAls}, near to the amorphous solidification 
transition the only linearly unstable band of $\pop$ is the long-wavelength 
band of the isotropic component
$\pop(\hk)$ (i.e.,~$\pop(\hk;\hl,\hm)\vert_{\hl=\hm=\hz}$ having sufficiently small 
(replicated) wave vectors, specifically,~$\hk^{2} < 2 \cp$).  
Thus, the basic process 
occurring at the transition is the acquisition of a nonzero value by the 
unstable components of the order parameter, which in turn perturb the stable 
components away from their zero values. The stable components include 
both the band of the isotropic component for $\hk^{2} > 2 \cp$ and 
the anisotropic components for all values of $\hk$.
As our aim is to calculate the leading contributions 
to each of the components of $\pop$ at small positive $\cp$, 
we may, as a first step, proceed by 
computing the self-consistent value of $\pop(\hk)$, neglecting the feedback 
on its value coming from the nonzero values that it induces in the  (stable)
anisotropic components [i.e.,~$\pop(\hk;\hl,\hm)\vert_{\hl\ne\hz}$]. 
Although the $\hk^{2} > 2 \cp$ band of the isotropic component $\pop(\hk)$ is 
linearly stable, it is necessary to treat it self-consistently together
with the unstable band, owing to the 
fact that they constitute a continuum (see Fig.~\ref{FIG:rail_roa}),
and therefore the stable band includes 
elements with an arbitrarily large ``susceptibility'' to perturbations caused by
their couplings to the elements of the linearly unstable band 
[i.e., $\pop(\hk)$ for $\hk^2 < 2 \cp$].
 
We therefore consider the \sce\ for $\pop(\hk;\hl,\hm)$, Eq.~(\ref{EQ:SCEexp}), 
for isotropic external arguments (i.e.,~$\hl=\hm=\hz$), and 
ignore the effects of all anisotropic components on the right hand side.  
We thus arrive at the closed, nonlinear 
\sce\ for $\pop(\hk_{0})$: 
\begin{equation}
\pop(\hk_{0})=
\left(
1-{\cp\over{3}}-{\hk^{2}\over{6}}
\right)
\pop(\hk_{0})
+{\cp\over{3}}
{\sqrt{4\pi} \over V^{n}}
\sum\nolimits_{\hk_{1}}
\pop(\hk_{1})
\pop(\hk_{0}-\hk_{1}).
\label{EQ:sceIsoUnsc}
\end{equation}
Precisely this equation emerges in the context of randomly crosslinked 
macromolecular networks, from both semi-microscopic and Landau-type 
approaches \cite{REF:cast,REF:review}.  In that context, the order parameter has 
only isotropic components, in contrast with the present context.  
To solve Eq.~(\ref{EQ:sceIsoUnsc}) we invoke the hypothesis 
for $\pop(\hk_{0})$ discussed in Sec.~\ref{SEC:locchardist}, 
viz., a parametrization in terms of the normalized probability distribution of
localization lengths $p(\tau)$, along with the $\delta_{\tilde{\bf k},{\bf 0}}$
factor, enforcing MTI of the solution: 
\begin{equation}
\pop(\hk_{0})=
{\delta_{\tilde{\bf k},{\bf 0}}\over{\sqrt{4\pi}}}
\int_{0}^{\infty}d\tau\,p(\tau)\,{\rm e}^{-\hk^{2}/2\tau}.
\label{EQ:POPisoAnsatz}
\end{equation}
As is shown in Ref.~\cite{REF:review}, this leads to the following
nonlinear integro-differential \sce\ for $p(\tau)$:
\begin{equation}
{\tau^{2}\over{2}}
{dp\over{d\tau}}
=\left({\cp\over{2}}-\tau\right)p(\tau)
-{\cp\over{2}}\int_{0}^{\tau}d\tau'\,p(\tau')\,p(\tau-\tau'),
\label{EQ:POPisoIDEuns}
\end{equation}%
i.e., precisely the equation found in Refs.~\cite{REF:cast,REF:review}.
By making the rescalings
\begin{mathletters}
\begin{eqnarray}
\tau   &\rightarrow&\theta\equiv2\tau/\cp, 
\\
p(\tau)&\rightarrow&\pi(\theta)\equiv{\cp\over{2}}p(\tau), 
\end{eqnarray}%
\end{mathletters}%
as in Refs.~\cite{REF:cast,REF:review}, we determine that the universal scaling 
function $\pi(\theta)$ satisfies the nonlinear integro-differential equation
\begin{equation}
{\theta^{2}\over{2}}
{d\pi\over{d\theta}}
=\left(1-\theta\right)\pi(\theta)
-\int_{0}^{\theta}d\theta'\,\pi(\theta')\,\pi(\theta-\theta'), 
\label{EQ:POPisoIDEsca}
\end{equation}
together with the normalization condition 
$\int_{0}^{\infty}d\theta\,\pi(\theta)=1$. 
The resulting scaled distribution $\pi(\theta)$ can be obtained numerically, 
and the result is given in Refs.~\cite{REF:cast,REF:review}.
(The prediction for this universal scaling function is compared with 
results from numerical simulations in Ref.~\cite{REF:Peng}.)\thinspace\ 
Thus we have obtained the isotropic component $\pop(\hk)$ 
of the order parameter $\pop(\hk;\hl,\hm)$ in the vicinity of the 
transition.  As we have discussed in Sec.~\ref{SEC:detector}, we are thus 
in possession of statistical information concerning the spatial 
localization of particles, regardless of the angular localization 
of the orbitals.
\subsection{Self-consistency equation: Anisotropic sectors}
\label{SEC:anisosector}
We now turn to the task of calculating the leading-order 
contributions to the {\it anisotropic} components of $\pop(\hk;\hl,\hm)$ 
[i.e.,~$\pop(\hk;\hl,\hm)\vert_{\hl\ne\hz}$].  We remind the reader 
that the anisotropic components are all linearly stable near the transition, 
and not merely infinitesimally so (i.e.,~linear stability analysis indicates that 
none of these anisotropic components even become marginally stable at the 
transition). Note the contrast with the stable band of the isotropic component
[$\pop(\hk)$ 
for $\hk^2 > 2 \cp$], which, though stable, do include marginally stable components
(i.e.,~components of arbitrarily small ``mass'').
Unlike the stable band of the isotropic component the 
(also stable) anisotropic components are
separated by a ``gap'' from the unstable components, owing to the discreteness 
of the external variable $\hl$, the components of which take on integer values only, 
in contrast with the components of $\hk$, which are continuous (in the thermodynamic
limit). It is this ``gap'' that allows us to obtain the stable anisotropic 
components by means of perturbation theory, which we could not use to solve for 
the stable band of the isotropic component (see Fig.~\ref{FIG:rail_roa}).
\subsubsection{Anisotropic sector: Angular momentum 1 in one replica channel}
\label{SEC:aniso0100}
Rather than begin with generalities, we first consider the lowest 
angular momentum sector $\{\pop(\hk;\hl,\hm)\}_{\hl^{2}=2}$ 
(i.e., the collection of order-parameter components $\pop(\hk;\hl,\hm)$ such 
that $\hl^2 \equiv \sum_{\alpha} \ell^{\alpha}(\ell^{\alpha}+1) = 2$. In this 
case $\hl$ is some permutation of the form $(0,1,0,0,\ldots,0)$. 
We therefore consider the \sce\ for $\pop\vtripW{1}(\hk)$, 
i.e., Eq.~(\ref{EQ:SCEexp}) for the anisotropic external argument 
\begin{equation}
\bordermatrix
{{\rm replica} &0 \phantom{,} &1 \phantom{,} &\ldots \phantom{,}
		&\alpha_{1} \phantom{,} &\alpha_{1}+1 \phantom{,}
		&\ldots \phantom{,} &n\cr
	    	\hl&0,&0,&\ldots,&1,        &0,          &\ldots,&0\cr
	        \hm&0,&0,&\ldots,&m_{1},    &0,          &\ldots,&0}, 
\label{EQ:WRAdef}
\end{equation}
and arbitrary $\hk$, which, as we establish in App.~\ref{APP:ASCE0100}, reads 
\begin{equation}
\pop\vtripW{1}(\hk)=
	\bfpc{0}\left(
 -\pop\vtripW{1}(\hk)
+{i\over{\sqrt{3}}}
 \pop(\hk_{1})\,\sck{1}
-q{\sqrt{4\pi}\over{V^{n}}}
\sum\nolimits_{\hk_{1}}
 \pop(\hk-\hk_{1})\,
 \pop\vtripW{1}(\hk_{1})
	\right).
\label{EQ:WRAsce}
\end{equation}
The symbol $\sck{}$ denotes the complex conjugate of the $m$-th
spherical tensor component of the vector ${\bf k}^{\alpha}$ 
spherical component of the vector ${\bf k}^{\alpha}$ [see 
App.~\ref{APP:ingred}, Eq.~(\ref{EQ:SCKdef}), for the definition].
The parameter $\bfpc{0}$ encodes physical information arising from 
orbital-orbital correlations of a single particle and is defined in
App.~\ref{APP:BFPC}. Specifically, $\bfpc{0}$ depends on the $\ell=1$ 
value of the free-particle two-orbital orientation correlator 
$\langle Y_{\ell m}^{\ast}({\bf s}_{a})\,
         Y_{\ell' m'}({\bf s}_{a'})\rangle_{1,1}$.  
The permutation symmetry among the $A$ orbitals and rotational invariance 
of the joint probability distribution of their orientations 
restrict this correlator to have the form  
\begin{equation}
\Big\langle Y_{\ell m}^{\ast}({\bf s}_{a})\,
        Y_{\ell' m'}({\bf s}_{a'})\Big\rangle_{1,1}
={1\over{4\pi}}\delta_{\ell,\ell'}\,\delta_{m,m'}
 \Big(\delta_{a,a'}+(1-\delta_{a,a'})\,\co{\ell}\Big), 
\label{EQ:BareCor}
\end{equation}
characterized by the parameters $\{\co{\ell}\}_{\ell=1}^{\infty}$ 
(with $\co{0}=1$, by normalization), introduced in Sec.~\ref{SEC:elements}.
As for the issue of what terms have been omitted in arriving at 
Eq.~(\ref{EQ:WRAsce}), we are concerned only with the leading-order values 
of the components of $\pop$, and therefore, here and elsewhere, shall 
omit terms that do not alter leading-order values.  In particular, 
in arriving at Eq.~(\ref{EQ:WRAsce}) from Eq.~(\ref{EQ:SCEexp}) we have 
neglected all components of $\pop$ having $\hl^{2}\ge 2$.  Such terms are 
sufficiently small that feedback from them would not alter the 
leading-order value of $\pop\vtripW{1}(\hk)$.  We shall verify 
the internal consistency of this assumption below.  In addition, 
we have only kept terms carrying sufficiently few powers of 
components of $\hk$.  As the characteristic value of $\hk^{2}$ 
is of order $\cp$, higher powers render terms subdominant. 
The steps leading from Eq.~(\ref{EQ:SCEexp}) to Eq.~(\ref{EQ:WRAsce}) 
are explained in detail in App.~\ref{APP:ASCE0100}.

To solve Eq.~(\ref{EQ:WRAsce}) we rewrite it in the form of a 
(Type~II inhomogeneous) integral equation by moving the first term 
on the right hand side to the left, regarding the second term as 
a known inhomogeneity, and the third term as a perturbation.  We then 
solve this equation iteratively, thus obtaining
\begin{equation}
\pop\vtripW{1}(\hk)=
{i\over{\sqrt{3}}}
\bfpc{1}\,
\pop(\hk)\,\sck{1},  
\label{EQ:WRAanswer}
\end{equation}
where the parameter $\bfpc{1}$ depends on $A$ and $\co{1}$ 
(see App.~\ref{APP:BFPC}).
In fact, for this particular component of $\pop$ our procedure 
merely amounts to truncating the Born series after the {\it zeroth\/} 
order (i.e.,~effectively ignoring the perturbation altogether), 
and solving the resulting algebraic equation.  However, for 
certain other components it turns out to be necessary to 
retain the first-order term, for reasons that we shall explain below. 

As we have discussed in Sec.~\ref{SEC:detector}, the result that 
we have just obtained about the value of the order-parameter component 
$\pop\vtripW{1}(\hk)$ yields statistical information concerning the 
variations, across the system, of the strength of the correlations 
between the thermal fluctuations of the positions of the localized particles 
and the thermal fluctuations of the orientations of their orbitals. We will
discuss this information in more detail in Sec.~\ref{SEC:PIlowaniso}
\subsubsection{Anisotropic sector: Angular momentum 1 in two replica channels}
\label{SEC:aniso0110}
Having obtained the leading-order contributions to the 
order-parameter components $\pop(\hk)$ and $\pop\vtripW{1}(\hk)$ 
[i.e., the isotropic (largest) and anisotropic (next largest) components], 
we now address the component of $\pop$ corresponding to 
\begin{equation}
\bordermatrix{{\rm replica}&0 \phantom{,} &1 \phantom{,} &\ldots \phantom{,} 
&\alpha_{1} \phantom{,} &\alpha_{1}+1 \phantom{,}&\ldots \phantom{,} &n\cr
	        \hl&0,&0,&\ldots,&1,        &0,          &\ldots,&0\cr
	        \hm&0,&0,&\ldots,&m_{1},    &0,          &\ldots,&0}, 
\label{EQ:TRAdef}
\end{equation}
which we denote by $\pop\vtripW{1}\vtripW{2}(\hk)$. 
The motivation for examining this component is that, in contrast to 
the components $\pop(\hk)$ and $\pop\vtripW{1}(\hk)$, the limit 
$\hk\rightarrow {\bfhz}$ of this component provides information purely about 
the orientational localization of the orbitals, independent of the 
positional localization properties, as we shall discuss in more detail 
in Sec.~\ref{SEC:PIlowaniso}.   

As shown in App.~\ref{APP:ASCE0110}, 
the retention of all terms that give rise to leading-order 
contributions to $\pop\vtripW{1}\vtripW{2}(\hk)$ 
leads to the following \sce\ for this component: 
\begin{eqnarray}
\pop\vtripW{1}\vtripW{2}(\hk)
&=&
 \bfpc{2}\pop\vtripW{1}\vtripW{2}(\hk)
-{1\over{3}}\bfpc{3}
 \sck{1}\,\sck{2}\,\pop(\hk)
+q\bfpc{2}{\sqrt{4\pi}\over{2V^{n}}} \sum\nolimits_{\hk_{1}}
 \pop(\hk-\hk_{1})\,\pop\vtripW{1}\vtripW{2}(\hk_{1})
\nonumber
\\
&&
\qquad\qquad\qquad\qquad
+q\bfpc{2}{\sqrt{4\pi}\over{V^{n}}}\sum\nolimits_{\hk_{1}}
 \pop\vtripW{1}(\hk-\hk_{1})\,\pop\vtripW{2}(\hk_{1}),
\label{EQ:TRAsce}
\end{eqnarray}%
where the parameters $\bfpc{2}$ and $\bfpc{3}$ depend on $A$ and $\co{1}$ 
(see App.~\ref{APP:BFPC}).  On the basis of the transformation 
properties of this equation under common rotations of the replicas
(and bearing in mind the coupling between positional and orientational 
degrees of freedom), we propose that the solution has the form 
\begin{equation}
\pop\vtripW{1}\vtripW{2}(\hk)=
 (-1)^{m_{1}}\,\delta_{m_{1}+m_{2},0}\,\iop^{(1)}(\hk^{2})
+\sck{1}\,\sck{2}\,\iop^{(2)}(\hk^{2}), 
\label{EQ:TRApropose}
\end{equation}
parametrized in terms of the two as-yet unknown functions
$\iop^{(1)}$ and $\iop^{(2)}$, which each depend only on $\hk^{2}$.
By inserting this proposed form into Eq.~(\ref{EQ:TRAsce}), contracting 
(on the indices $m_{1}$ and $m_{2}$),
first with $ (-1)^{m_{1}}\,\delta_{m_{1}+m_{2},0}$ and then
with $\sck{1}\,\sck{2}$, and considering the limit 
${\bf k}^{\alpha^{1}}={\bf k}^{\alpha^{2}}\rightarrow{\bf 0}$ 
(with $\hk^{2}$ fixed and arbitrary), and retaining only terms 
that contribute to the leading-order value of 
$\pop\vtripW{1}\vtripW{2}(\hk)$,
we arrive at the pair of coupled 
(Type~II inhomogeneous) integral equations
\begin{mathletters}
\begin{eqnarray}
\bfpc{4}\,\iop^{(2)}(\hk^{2})
&=&-{1\over{3}}\,{\bfpc{3}\over{\bfpc{2}}}\,\pop(\hk),
\label{EQ:isoTwo}
\\
\bfpc{4}\,\iop^{(1)}(\hk^{2})
&=& 
{q\,\sqrt{4\pi} \over {V^{n}}}
\sum\nolimits_{\hk_{1}}\pop(\hk-\hk_{1})\,\iop^{(1)}(\hk_{1}^{2})
+q\,{\sqrt{4\pi} \over {V^n}}
\bigg\{ 
 \sum\nolimits_{\hk_{1}}\pop\vtripW{1}(\hk-\hk_{1})\,\pop\vtripW{2}(\hk_{1})
\bigg\}_{\delta}
\nonumber
\\
&&
\qquad\qquad\qquad\qquad
+q\,{\sqrt{4\pi} \over {V^n}}
\bigg\{
\sum\nolimits_{\hk_{1}}
	\sckW{1}\,\sckW{2}\,
	\iop^{(2)}(\hk_{1}^{2})\,\pop(\hk-\hk_{1})
	\bigg\}_{\delta},
\label{EQ:isoOne}
\end{eqnarray}%
\end{mathletters}%
where the parameter $\bfpc{4}$ depends on $A$ and $\co{1}$ 
(see App.~\ref{APP:BFPC}). 
The symbol $\liso\cdots\riso$ denotes the result of extracting 
from the quantity inside the braces the coefficient of the term 
proportional to the 
isotropic rank-2 spherical tensor $(-1)^{m_{1}}\delta_{m_{1}+m_{2},0}$. 
[To extract this part, we take the limit 
${\bf k}^{\alpha^{1}}={\bf k}^{\alpha^{2}}\rightarrow{\bf 0}$ 
with $\hk^{2}$ fixed and arbitrary, and contract with 
$(-1)^{m_{1}}\delta_{m_{1}+m_{2},0}/3$.]\thinspace\ To find the leading 
contributions to $\iop^{(1)}$ and $\iop^{(2)}$ we first read off 
the value of the latter from Eq.~(\ref{EQ:isoTwo}).  We then 
use this result to eliminate $\iop^{(2)}$ from Eq.~(\ref{EQ:isoOne}), 
observing that we can omit the first term on the right hand side 
of Eq.~(\ref{EQ:isoOne}), and perform the remaining summations 
to arrive at the results
\begin{mathletters}
\begin{eqnarray}
\iop^{(2)}(\hk^{2})
&=&
-{1\over{3}}\,\bfpc{5}
 {1\over{\sqrt{4\pi}}}
 \int_{0}^{\infty}d\theta\,
 \pi(\theta)\,{\rm e}^{-\hk^{2}/\cp\theta}, 
\label{EQ:ansTwo}
\\
\iop^{(1)}(\hk^{2})
&=& 
{\bfpc{6}\over{\sqrt{4\pi}}}{\cp^2\over{4}}\,
	\int_{0}^{\infty}d\theta\,
	\piconv(\theta)\,
{\rm e}^{-\hk^{2}/\cp\theta},\,\, {\rm where}
\label{EQ:ansOne}
\\	
\theta \piconv(\theta) &\equiv&
\int_{0}^{\infty}
d\theta_{1}\,\pi(\theta_{1})\,
	d\theta_{2}\,\pi(\theta_{2})\,
	\theta_{1}\theta_{2}
	\delta\big(\theta-(\theta_{1}+\theta_{2})\big)
	=
	\Big(
	\big(\theta \, \pi(\theta)\big)\ast
	\big(\theta \, \pi(\theta) \big)
	\Big) (\theta),
\end{eqnarray}%
\end{mathletters}%
where the symbol $\ast$ represents Laplace convolution.  
Note that, as can be anticipated from our perturbative approach to obtaining
the anisotropic components, the results for $\iop^{(1)}$ and $\iop^{(2)}$
are constructed from the universal scaling function $\pi(\theta)$.
\subsubsection{Self-consistency equation: General case}
\label{SEC:anisoGeneral}
Having obtained the isotropic component of $\pop$, as well as the two lowest 
angular momentum anisotropic components, we now address the task of establishing 
the general form of $\pop$, along with an algorithm for obtaining the 
leading-order contributions to $\pop$ for 
arbitrary values of its arguments. We begin by proposing the following structure 
for the general form of the leading-order contribution to $\pop(\hk;\hl,\hm)$:
\begin{equation}
\pop(\hk;\hl,\hm) \approx \poly_{\hl,\hm}(\hk)\, \pop(\hk) + 
\cp^{1+ \frac{1}{2} \tilde{\ell} }\,
\tens_{\hl,\hm}\,
\int_{0}^{\infty} d\theta\,\pi_{\hl,\hm}(\theta)\,{\rm e}^{-\hk^{2}/\cp\theta}.
\label{EQ:POPstruc}
\end{equation}
Here $\poly_{\hl,\hm}(\hk)$ is a certain homogeneous 
polynomial in $k^{(\ell)}_{m}$ in which 
all terms are of order $\tilde{\ell}$ in $k$. (We remind the reader that
$k^{(\ell)}_{m}$ stands for $\sqrt{(2 \ell + 1)/4 \pi}\,
k^{\ell}\, Y_{l m}\left({\bf k}/k\right)$, and is thus of 
order $\ell$ in $k$.)\thinspace In addition, $\tens_{\hl,\hm}$ is a certain
rotationally-invariant tensor depending only on $\hl$ and $\hm$, and 
$\pi_{\hl,\hm}(\theta)$ is a distribution. All the unknown ingredients
[$\poly_{\hl,\hm}(\hk)$, $\tens_{\hl,\hm}$, and $\pi_{\hl,\hm}(\theta)$]
will be obtained below.

To illustrate these notions with a concrete example we note that 
$\pop\vtripW{1}\vtripW{2}(\hk)$, constructed in Sec.~\ref{SEC:aniso0110}
has such a form, with
\begin{mathletters}
\begin{eqnarray}
\poly_{\hl,\hm}(\hk)\vert_{\hl = (0,\ldots,0,1,1,0,\ldots,0)} &=& 
-{1\over{3}}\,\bfpc{5}\,\sck{1}\,\sck{2},
\\
\tens_{\hl,\hm}\vert_{\hl = (0,\ldots,0,1,1,0,\ldots,0)} &=& 
(-1)^{m_1} \delta_{m_1+m_2,0},
\\
\pi_{\hl,\hm}(\theta)\vert_{\hl = (0,\ldots,0,1,1,0,\ldots,0)} &=&
{\bfpc{6}\over{4 \sqrt{4\pi}}}\,
\piconv(\theta),
\end{eqnarray}%
\end{mathletters}%
where $\hl = (0,\ldots,0,1,1,0,\ldots,0)$ indicates that 
$\ell^{\alpha_1} = \ell^{\alpha_2} = 1$, with $\ell^{\alpha}=0$ in all other 
replicas. Note that 
$\poly_{\hl,\hm}(\hk)\vert_{\hl = (0,\ldots,0,1,1,0,\ldots,0)}$ is of order
${\tilde \ell} = 2$ in components of $\hk$.

To show that the leading-order contribution to the general component
$\pop(\hk;\hl,\hm)$ in the vicinity of the amorphous solidification transition
does indeed have the form~(\ref{EQ:POPstruc}),
for all values of $\hl$, $\hm$, and $\hk$,
we proceed by full mathematical induction on the multi-index 
$\hl$ \cite{REF:induction}. We note that $\pop(\hk)$ has 
this form with $\poly_{\hz,\hz}(\hk) = 1$ and $\pi_{\hz,\hz}(\theta) = 0$, thus
establishing the {\it base} of induction. To establish the 
{\it step} of induction we assume
that $\pop(\hk;\hl,\hm)$ has the form~(\ref{EQ:POPstruc}) for all values 
$\hl < \hl_0$ (by which we mean $\ell^{\alpha} \le \ell_0^{\alpha}$ for all $\alpha$,
and $\ell^{\alpha} < \ell_0^{\alpha}$ for at least one $\alpha$). We then examine
the \sce\ for $\pop(\hk_0;\hl_0,\hm_0)$:
\begin{eqnarray}
\pop(\hk_{0};\hl_{0},\hm_{0}) &=&
(1-q)
{4\pi\mu^{2} \over {V^{n}}}\,
\sum_{\hk_{1},\hl_{1},\hm_{1}}
\pop(\hk_{1};\hl_{1},\hm_{1})\,
{1\over{A^{2}}}
\sum_{a_{0},a_{1}=1}^{A}
\left\langle
{\rm e}^{-i\hk_{0}\cdot\hat{\bf c}}\,
\hy_{\hl_0 \hm_0}^{\ast}(\hs_{a_0})\,
{\rm e}^{i\hk_{1}\cdot\hat{\bf c}}\,
{\rm e}^{i\hk_{1}\cdot\hat{\bf s}_{a_{1}}}\,
\hy_{\hl_1 \hm_1}(-\hs_{a_1})
\right\rangle_{1,n+1}
\nonumber
\\
&&\qquad
+\frac{1}{2}
\left({4\pi\mu^{2}\over{V^{n}}}\right)^{2}\,q
\sum_{\hk_{1},\hl_{1},\hm_{1}}
\sum_{\hk_{2},\hl_{2},\hm_{2}}
\pop(\hk_{1};\hl_{1},\hm_{1})\,
\pop(\hk_{2};\hl_{2},\hm_{2})
\nonumber
\\
&\times&
{1\over{A^{3}}}\sum_{a_{0},a_{1},a_{2}=1}^{A}
\left\langle
{\rm e}^{-i\hk_{0}\cdot\hat{\bf c}}\,
\hy_{\hl_0 \hm_0}^{\ast}(\hs_{a_0})\,
{\rm e}^{i\hk_{1}\cdot\hat{\bf c}}\,
{\rm e}^{i\hk_{1}\cdot\hat{\bf s}_{a_{1}}}\,
\hy_{\hl_1 \hm_1}(-\hs_{a_1})\,
{\rm e}^{i\hk_{2}\cdot\hat{\bf c}}\,
{\rm e}^{i\hk_{2}\cdot\hat{\bf s}_{a_{2}}}\,
\hy_{\hl_2 \hm_2}(-\hs_{a_2})
\right\rangle_{1,n+1}.
\label{EQ:SCEgen}%
\end{eqnarray}%
As we are are only concerned with leading-order contributions to $\pop$, we
truncate the $\hl$ sums in Eq.~(\ref{EQ:SCEgen}) so as to include only terms with
$\hl \le \hl_0$. We now examine the contributions to $\pop(\hk_0,\hl_0,\hm_0)$
coming from linear couplings to lower-angular momentum components of $\pop$ 
[i.e.,~the terms in the right-hand side of Eq.~(\ref{EQ:SCEgen}) that are
linear in $\pop$]. Consider the coupling to 
$\pop(\hk,\hl,\hm)$. By the translational invariance of the correlator, 
performing the $\hat{{\bf c}}$ average in 
Eq.~(\ref{EQ:SCEgen}) we establish that $\hk = \hk_0$. The remaining correlator 
factorizes on the replica index, becoming
\begin{equation}
\prod_{\alpha=0}^{n}\left\langle
Y_{\ell_{0}^{\alpha} m_{0}^{\alpha}}^{\ast}({\bf s}_{a_{0}}^{\alpha})\,
Y_{\ell^{\alpha} m^{\alpha}}(-{\bf s}_{a_{1}}^{\alpha})\,
{\rm e}^{i {\bf k}_0^{\alpha}\cdot\hat{\bf s}^{\alpha}_{a_{1}}}
\right\rangle_{1,n+1}.
\end{equation}
By introducing the Rayleigh-expansion, Eq.~(\ref{EQ:rayleigh}), for the factor
$\exp i {\bf k}_0^{\alpha}\cdot\hat{\bf s}^{\alpha}_{a_{1}}$ we see that 
this correlator is nonzero only for the terms of order $\ell'^{\alpha}$ 
in the Rayleigh expansion for which the angular momenta $\ell'^{\alpha}$, 
$\ell_0^{\alpha}$, and $\ell^{\alpha}$ can sum to angular momentum $0$. As we are 
only interested in the leading-order contributions to $\pop$, and the 
$\ell'^{\alpha}$ term in the Rayleigh expansion is of order 
${(k_0^{\alpha})}^{{\ell'}^{\alpha}}$, we need only keep the lowest angular-momentum
term (i.e., $\ell'^{\alpha} = \ell_0^{\alpha} - \ell^{\alpha}$).  Thus, from each
replica we shall pick up the factors
${(k_0^{\alpha})}_{m_0^{\alpha}-m^{\alpha}}^
{\left( \ell_0^{\alpha} - \ell^{\alpha}\right)}$, 
provided, of course, that
$\left\vert m_0^{\alpha}-m^{\alpha} \right\vert \le \ell_0^{\alpha} - \ell^{\alpha}$.
Together from all replicas,
these factors will give a factor, depending on the $\{k^{(\ell)}_{m}\}$, that
will be of order $\tilde{\ell_0} - \tilde{\ell}$ in $k$. To leading order, the term 
that corresponds to the coupling of $\pop(\hk_0,\hl_0,\hm_0)$ to $\pop(\hk,\hl,\hm)$ 
will thus become,
\begin{equation}
\left(\prod_{\alpha=0}^{n} 
{(k_0^{\alpha})}_{m_0^{\alpha}-m^{\alpha}}^
{\left(\ell_0^{\alpha} - \ell^{\alpha}\right)} 
C_{\hl_0,\hm_0;\hl,\hm}^{\alpha} \right)\,
\poly_{\hl,\hm}(\hk_0)\, \pop(\hk_0),
\end{equation}
where the $C_{\hl_0,\hm_0;\hl,\hm}^{\alpha}$ are constants, for which we cannot 
in general provide a closed-form result, but which we can, 
however, compute, should we decide to construct some
component of the order parameter explicitly, as we have, e.g., done for
$\pop\vtripW{1}\vtripW{2}$.

As $\prod_{\alpha} {(k_0^{\alpha})}^{\left(\ell_0^{\alpha} - \ell^{\alpha}\right) }
_{m_0^{\alpha}-m^{\alpha}}$ 
is of order $\tilde{\ell_0} - \tilde{\ell}$ in $k_0$, and as
$\poly_{\hl,\hm}(\hk_0)$ is, according to the inductive assumption, of order 
$\tilde{\ell}$, we see that this contribution is of order $\tilde{\ell_0}$ 
in $k_0$. Taking the term corresponding to $\hl=\hl_0, \hm=\hm_0$ over to the left-hand side, adding all
remaining terms, and making the definition
\begin{equation}
\poly_{\hl_0,\hm_0}(\hk_0) \equiv 
\sum_{\hl < \hl_0} \sum_{\hm} \left(\prod_{\alpha=0}^{n} 
C_{\hl_0,\hm_0;\hl,\hm}^{\alpha}\,
{(k_0^{\alpha})}_{m_0^{\alpha}-m^{\alpha}}^
{\left(\ell_0^{\alpha} - \ell^{\alpha}\right) } 
\right) \poly_{\hl,\hm}(\hk_0),
\end{equation}
we see that, to leading order in $k_0$, the linear contribution to 
$\pop(\hk_0;\hl_0,\hm_0)$ is indeed of the
form~(\ref{EQ:POPstruc}), with $\poly_{\hl_0,\hm_0}(\hk_0)$ being of the correct
order in $k_0$. As, due to the fact that $q$ is small near the transition, 
the linear contribution is of lower order in $\cp$ than the
quadratic terms for values of $\hk^2$ of order $\cp$, we have thus also 
established a recursive algorithm for determining $\poly_{\hl_0,\hm_0}(\hk_0)$. 
Illustrations of its use can be found in Apps.~\ref{APP:ASCE0100} 
and~\ref{APP:ASCE0110}.

We now examine the {\it quadratic} contribution to $\pop(\hk_0;\hl_0,\hm_0)$ 
in Eq.~(\ref{EQ:POPstruc}). 
This term is only the dominant one for $\hk_0^2 \ll \cp$ and, hence, we need 
only obtain it in the 
limit $\hk_0^2 \rightarrow 0$, which is equivalent to extracting from it 
the leading-order contribution to the part
proportional to the (${\bf k}$ - independent) 
rotationally-invariant tensor $\tens_{\hl_0,\hm_0}$. 
As we are only interested in the
leading-order behavior of $\pop$, in the summations over $\hl$ in the 
quadratic term of Eq.~(\ref{EQ:SCEgen}) we need only include the linear
contributions to $\pop(\hk;\hl,\hm)$ [i.e.,~for the purposes of computing the
quadratic term of $\pop(\hk_0;\hl_0,\hm_0)$ we can set
$\pop(\hk;\hl,\hm) = \poly_{\hl,\hm}(\hk)\, \pop(\hk)$ for $\hl \le \hl_0$].
We now study the quadratic coupling of $\pop(\hk_0;\hl_0,\hm_0)$ to lower 
angular-momentum components. In each replica $\alpha$ we have four sources 
of angular momentum: two ($\ell_1^{\alpha}$ and $\ell_2^{\alpha}$)
coming from the two components of $\pop$ [i.e.,
$\pop(\hk_1;\hl_1,\hm_1)$ and $\pop(\hk_2;\hl_2,\hm_2)$]; and two 
(${\ell'}_1^{\alpha}$ and ${\ell'}_2^{\alpha}$), one
coming from each of the Rayleigh expansions of the ``shift'' factors 
$\exp i\hk_{1}\cdot\hat{\bf s}_{a_{1}}$ and 
$\exp i\hk_{2}\cdot\hat{\bf s}_{a_{2}}$.  
As we are only interested in
the leading-order behavior of $\pop$, we need only consider the case when
$\ell_1^{\alpha} + \ell_2^{\alpha}+{\ell'_1}^{\alpha}+{\ell'}_2^{\alpha} =
\ell_0^{\alpha}$. Each source of angular momentum $\ell$ brings with it
a factor of $k^{\ell}$. Hence, multiplying all factors from all replicas 
together and assembling all terms in the summations we shall obtain
\begin{equation}
q  V^{-n} 
\bigg\{
\sum\nolimits_{\hk_1,\hk_2} {\cal F}(\hk_1,\hk_2)\,\pop(\hk_1)\,\pop(\hk_2)\, 
\delta_{\hk_1+\hk_2,\hk_0}\,
\delta_{{\tilde{\bf k}}_{1},{\bf 0}}\, 
\delta_{{\tilde{\bf k}}_{2},{\bf 0}}
\bigg\}_{\tens},
\end{equation}
where the operation $\{\cdots\}_{\tens}$ denotes the extraction, from $\cdots$,
of the part proportional to $\tens_{\hl_0,\hm_0}$, and where
${\cal F}$ is a polynomial function of $k_1$ and $k_2$ in which all terms
are of order $\tilde \ell_0$. Provided, that
$\hl_0$ and $\hm_0$ satisfy conditions for macroscopic rotational invariance 
(note, in particular, that MRI requires $\tilde{\ell_0}$ to be even and, hence, 
that the quadratic contribution vanishes for $\tilde{\ell_0}$ odd),
we can perform the summations,
indeed obtaining the claimed structure for the second-order term 
in Eq.~(\ref{EQ:POPstruc}). Thus, we have established 
a recursive procedure for obtaining $\pi_{\hl_0,\hm_0}(\theta)$.
Keeping in mind that typical values of $\hk_0^2$ are of order $\cp$, that 
$q=2/3 \cp$, and that $\pop(\hk)$ is of order unity we 
verify the scaling of the quadratic term
to be $\cp^{(\tilde{\ell_0}/2) + 1}$, thus completing the {\it step} of induction, 
viz. the second and final element of our proof.

In conclusion of our discussion of the solution of the order-parameter \sce\ we note, 
in passing, that all the assumptions that we have made in Sec.~\ref{SEC:strategy} 
regarding the scaling (with $\cp$) of various quantities are verified, {\it a posteriori\/}, 
by the solution that we have obtained, thus establishing the self-consistency of 
these assumptions.

\section{Physical information encoded in the order parameter}
\label{SEC:PhysInfo}
\subsection{Introduction}
\label{SEC:PIintro}
We have constructed the solution of the order-parameter self-consistency 
equation in the vicinity of the amorphous solidification transition, 
Eq.~(\ref{EQ:POPstruc}), and have obtained explicit solutions for the two 
lowest angular-momentum components of the order parameter, 
Eqs.~(\ref{EQ:WRAanswer},\ref{EQ:ansTwo},\ref{EQ:ansOne}). We are thus 
in possession of a range of statistical information about the amorphous 
solid state, this information being encoded in the order parameter, as 
discussed in Sec.~\ref{SEC:detector}.  In this section we will extract
some of this information explicitly, from the two lowest 
angular-momentum components of the order parameter, provide a strategy
for obtaining other statistical information about the system from
higher angular-momentum components, and discuss the scaling of the
order parameter with $\cp$ near the transition, as well as the 
implications of this scaling.

Our statistical diagnosis of the structure of the amorphous solid state
in the vicinity of the amorphous solidification transition is made in
terms of the moments of the joint distribution function $P$ that
collects together the localization characteristics of all the particles
in the sample and their orbitals, averaged over realizations of the
disorder. As we shall see, we are unable to construct the entire
distribution function, or even to construct a closed-form expression for
arbitrary moments. However, most of the useful information about the
system can be obtained from low moments of the distribution, resulting
from the two lowest angular-momentum components of the order parameter,
which we have obtained explicitly. As for the information encoded in the
higher angular-momentum components of the order parameter, we do not
extract it explicitly. We do, however, describe the kind of information
that could be obtained from them, as well as provide the general
procedure for doing so.

\subsection{Encoded information: 
Isotropic sector}
\label{SEC:PIisosector}
We begin by discussing the statistical information, encoded in the
isotropic component of the order parameter $\pop(\hk^{2})$. In
Sec.~\ref{SEC:isosector} we have obtained this component in the vicinity
of the amorphous solidification transition and, consequently, the
reduced distribution of (inverse square) localization lengths $p(\tau)$
associated with localized particles.  The fact that $p(\tau)$ takes on a
scaling form, and that the scaling function $\pi(\theta)$ has a well
defined peak with location and width both of order unity, allows us to
establish that (as $\cp\to 0$) $\tau$ scales as $\cp$ and, accordingly,
$\xi$ scales as $\cp^{-1/2}$. For example, any reasonable choice for a
characteristic value of $\xi$, say
$\int_{0}^{\infty}d\tau\,p(\tau)\,\tau^{-1/2}$, 
scales as $\cp^{-1/2}$.

In the present context, while the emergent distribution $p(\tau)$ is
found to have striking scaling properties, there is the suspicion that
changes in the details of the model will lead to changes at least in the
details of the scaling function $\pi(\theta)$ or, perhaps, in the
scaling property itself.  However, it has been found in the context of
vulcanized macromolecular matter that the scaling property, as well as
the precise form of the scaling function, are robust, universal features
of the mean-field theory, verified by independent computer simulations.
Moreover, this universality has been found to have its origins in the
symmetries of the appropriate Landau free energy and the divergence (at
the transition) of the characteristic localization length.

Being in possession of the entire distribution of localization lengths
provides us with a surprisingly rich characterization of the positional
aspects of the amorphous solid state.  It is striking that the
distribution is universal, not only across the macromolecular systems
where it was first found, but also in the present setting of vitreous
media.  Although we have obtained the distribution via analysis of a
specific semi-microscopic model of vitreous media, we anticipate that,
here too, the result will have a broader domain of applicability.
Moreover, given the emerging picture of orientational order as order
slaved to the underlying positional order, it is not surprising---and
indeed we shall demonstrate this point below---that all other
statistical descriptors of the amorphous solid state are also
constructed from the universal function $\pi(\theta)$.
\subsection{Encoded information: 
Low angular-momentum anisotropic sectors}
\label{SEC:PIlowaniso}
\subsubsection{Angular localization}
\label{SEC:PIlowanisoang}
We now discuss some of the specific physical information that can be 
obtained by examining our explicit solutions for the two lowest 
angular-momentum components of the order parameter, which we have 
obtained in Secs.~\ref{SEC:aniso0100} and \ref{SEC:aniso0110}. 
The first piece of information concerns the angular localization of 
the orbitals, without regard to the positional localization of the particles. 
As we recall from Sec.~\ref{SEC:detector}, such information is accessed via 
the order parameter for $\hk=\bfhz$, and is described by the distribution 
of the collection of characteristics $\{ \aaop_{\ell m} \}$. 
The lowest angular-momentum order-parameter component that provides access 
to this information is $\pop\vtripW{1}\vtripW{2}(\hk)$, the solution for 
which is given by Eq.~(\ref{EQ:TRApropose}).  
Evaluating at $\hk=\bfhz$ gives 
\begin{equation}
q\,\pop\vtripW{1}\vtripW{2}({\bfhz})= 
(-1)^{m_1}\delta_{m_1 + m_2 , 0}
{\bfpc{6}\over{\sqrt{4\pi}}}{\cp^3\over{6}}\,
\int_{0}^{\infty}d\theta\,\piconv(\theta). 
\label{EQ:TRAkz}
\end{equation}
Now, recalling the interpretation of the order parameter 
developed in Sec.~\ref{SEC:detector}, we arrive 
\begin{eqnarray}
q\,\int d\tau\,d\{\aaop\}\,{\cal D}\{\pacf\}\, 
P(\tau,\{\aaop\},\{\pacf\})\,\,
\aaop_{1,m_1}\,\aaop_{1,m_2} 
&=&
\left[
{1\over{N}}\sum_{j=1}^{N}
{1\over{A}}\sum_{a=1}^{A}
\Big\langle Y_{1 m_1}^{\ast}({\bf s}_{j,a})\Big\rangle\,
\Big\langle Y_{1 m_2}^{\ast}({\bf s}_{j,a})\Big\rangle
\right]
\nonumber
\\
&=&(-1)^{m_1}\delta_{m_1+m_2,0}
{\bfpc{6}\over{\sqrt{4\pi}}}{\cp^3\over{6}}
\int_{0}^{\infty}d\theta\,\piconv(\theta). 
\label{EQ:AngInfo0110}
\end{eqnarray}%
Notice that this characteristic of the angular localization of the 
orbitals is essentially the order parameter traditionally used to describe 
the directional localization of magnetic moments in the spin-glass state. 
In fact, if we recall the spherical-tensor decomposition of the scalar 
product of two unit vectors, 
${\bf s}_1\cdot {\bf s}_2 = 
\sum_m (-1)^m Y_{1 m}({\bf s}_1) Y_{1 -m}({\bf s}_2)$, 
then by appropriately contracting Eq.~(\ref{EQ:AngInfo0110}) over 
$m_1$ and $m_2$ we obtain the familiar characterization of directional 
localization: 
\begin{equation}
\left[
{1\over{N}}\sum_{j=1}^{N}
{1\over{A}}\sum_{a=1}^{A}
\langle{\bf s}_{j,a}\rangle\cdot
\langle{\bf s}_{j,a}\rangle
\right] 
={3\,\bfpc{6}\over{\sqrt{4\pi}}}{\cp^3\over{6}}\,
	\int_{0}^{\infty}d\theta\,\piconv(\theta). 
\label{EQ:spinglassop}
\end{equation}

\subsubsection{Angle-position fluctuation correlations}
\label{SEC:PIlowanisocorr}
Further specific physical information concerns the degree to which the 
thermal fluctuations in the orbital orientations are correlated with the 
thermal fluctuations of the particle positions.  As discussed in 
Sec.~\ref{SEC:detector}, to extract this information, 
which is encoded in the distribution of the collection of functions 
$\{\pacf_{\ell m}(\hk)\}$, we examine the order-parameter component 
$\pop\vtripW{1}(\hk)$:
\begin{eqnarray}
q\,\pop\vtripW{1}(\hk)
&=&
q\,\int d\tau\,d\{\aaop\}\,{\cal D}\{\pacf\}\, 
P(\tau,\{\aaop\},\{\pacf\})\,
\pacf_{1,m_1}({\bf k}^{\alpha_{1}})\,
{\rm e}^{-\hk^{2}/2\tau}
\nonumber
\\
&=&
\Big[
{1\over{N}}\sum_{j=1}^{N}
{1\over{A}}\sum_{a=1}^{A}
	\Big\langle
\Big(
{\rm e}^{-i{\bf k}^{\alpha_{1}}\cdot({\bf c}_{j}-{\bfmu}_{j})}-
\big\langle
{\rm e}^{-i{\bf k}^{\alpha_{1}}\cdot({\bf c}_{j}-{\bfmu}_{j})}
\big\rangle
\Big)
\Big(
Y_{1 m_1}^{\ast}({\bf s}_{j,a})- 
\big\langle 
Y_{1 m_1}^{\ast}({\bf s}_{j,a})
\big\rangle
\Big) 
\Big\rangle
\Big] 
\nonumber
\\
&=&
{2\cp\over {3}}
{i\bfpc{1}\over{\sqrt{3}}}\,\sck{1}\, 
{\delta_{\tilde{\bf k},{\bf 0}}\over{\sqrt{4\pi}}}
\int_{0}^{\infty}d\theta\,\pi(\theta)\,{\rm e}^{-\hk^{2}/ \cp \theta}.
\label{EQ:CorrInfo0100}
\end{eqnarray}%
By considering the derivative with respect to ${\bf k}^{\alpha_{1}}$, 
taking the limit $\hk\to{\bfhz}$, and contracting, we obtain
\begin{equation}
\left[
{1\over{N}}\sum_{j=1}^{N}
{1\over{A}}\sum_{a=1}^{A}
\Big\langle
\big({\bf c}_{j}- \langle{\bf c}_{j}\rangle\big)
\cdot
\big({\bf s}_{j,a} - \langle{\bf s}_{j,a}\rangle\big)
\Big\rangle 
\right]
\approx
-{\cp \over{\sqrt{3\pi}}}\,\bfpc{1}
\end{equation}
characterizing the anticorrelation of the thermal
orientation-fluctuations of orbitals with the thermal
position-fluctuations of the particles to which the orbitals are
attached. 
\subsection{Encoded information: 
Higher angular-momentum anisotropic sectors}
\label{SEC:PIhigheraniso}
In the previous two subsections we have obtained explicit physical 
information about the amorphous solid state in the vicinity of the 
solidification transition from the three lowest angular-momentum 
components of the order parameter.  We now discuss the generalization 
of the ideas that we have used in extracting that information.

Take, for instance, the task of obtaining information about the 
orientational localization of the orbitals.  Although the spin glass order-parameter--like quantity 
$[\langle{\bf s}\rangle\cdot\langle{\bf s}\rangle]$, 
constructed from 
$\big[\big\langle Y_{1 m_1}({\bf s})\big\rangle\,
\big\langle Y_{1 m_2} ({\bf s}) \big\rangle \big]$, 
provides the simplest characterization of the orientational 
localization of the orbitals, one could also study its higher 
angular momentum analogues, 
such as, e.g., 
$\big[\big\langle Y_{2 m_1}({\bf s})\big\rangle\,
      \big\langle Y_{2 m_2}({\bf s})\big\rangle\big]$ 
or, in general,
$\big[\prod_{\alpha}\left\langle Y_{\ell^{\alpha} m^{\alpha}}({\bf s}) 
\right\rangle\big]$. 
Such quantities could be useful if, e.g., one wished to study the 
properties of the formed covalent bonds separately from those of the 
unbonded orbitals.  Such quantities are obtained as appropriate moments 
of the joint probability distribution $P$ which, in turn, are extracted 
from the corresponding components of the order parameter.  Specifically,  
$\pop(\hz;\hl,\hm)$ yields 
$\big[\prod_{\alpha}\left\langle Y_{\ell^{\alpha} m^{\alpha}}({\bf s}) 
\right\rangle\big]$.  In Sec.~\ref{SEC:anisoGeneral} we have established 
an algorithm that allows us to obtain the leading-order contributions to 
all components of the order parameter.  Accordingly, we also have an 
algorithm for obtaining a large class~\cite{REF:moments} of the moments 
of the joint probability distribution $P$, including 
$\big[\prod_{\alpha}\left\langle Y_{\ell^{\alpha} m^{\alpha}}({\bf s}) 
\right\rangle\big]$ for arbitrary $\hl$ and $\hm$, via  
\begin{equation}
\left[
{1\over{N}}\sum_{j=1}^{N}
{1\over{A}}\sum_{a=1}^{A}
\prod_{\alpha=0}^{n}\left\langle Y_{\ell^{\alpha} m^{\alpha}}({\bf s})\right\rangle
\right]
=q\int d\tau\,d\{\aaop\}\,{\cal D}\{\pacf\}\, 
P(\tau,\{\aaop\},\{\pacf\})\,
\hat{\aaop}_{\hl \hm} 
= q\,\pop(\hz;\hl,\hm),
\label{EQ:AngInfoGen}
\end{equation}
where $\hat{\aaop}_{\hl \hm}$ was used to denote
$\prod_{\alpha} \aaop_{\ell^{\alpha} m^{\alpha}}$.

Similarly, higher-order characterizations of the thermal 
orientation-position correlations, i.e., the moments 
\begin{equation}
q\,\int d \tau \, d \{\aaop\} \, {\cal D} \{\pacf\} \, 
P(\tau,\{\aaop\},\{\pacf\})\,
\hat{\pacf}_{\hl \hm} (\hk)
 \exp(-\hk^2 / 2 \tau), 
\end{equation}
can be obtained from $q\,\pop(\hk;\hl,\hm)$.
[Note that $\hat{\pacf}_{\hl \hm} (\hk)$ denotes
$\prod_{\alpha} \pacf_{\ell^{\alpha} m^{\alpha}}({\bf k}^{\alpha})$.]
\thinspace For $\tilde{\ell}$ odd these moments are simply equal to 
$q\,\pop(\hk;\hl,\hm)$. 
For $\tilde{\ell}$ even, they are equal to 
$q\,\pop(\hk;\hl,\hm)-q\,\pop({\bfhz};\hl,\hm)$.  
Then, to find such characterizations, just as we did in the previous 
section, one needs to extract the coefficient of 
$\left({\bf k}^{\alpha}\right)^{\ell^{\alpha}}_{m^{\alpha}}$, and 
pass to the limit $\hk\to{\bfhz}$.  For example, by applying this 
procedure to $\pop\vtripW{1}\vtripW{2}(\hk)$ we would obtain 
\begin{equation}
\bigg[
{1\over{N}}\sum_{j=1}^{N}
{1\over{A}}\sum_{a=1}^{A}
\Big(
\Big\langle  
\big(
{\bf c}_{j}-\langle{\bf c}_{j}\rangle
\big)\cdot\big(
{\bf s}_{j,a} - \langle{\bf s}_{j,a}\rangle
\big)
\Big\rangle
\Big)^2
\bigg] 
\approx
{\cp\over{\sqrt{4 \pi}}}
\,\bfpc{5}. 
\end{equation}
\subsection{Scaling and its indications}
\label{SEC:PIscaling}
We now address the manner in which various physical quantities reflecting 
the orientational ordering scale with $\cp$.  To do this we examine the 
scaling of various components of the order parameter, established in 
Sec.~\ref{SEC:anisoGeneral}.  Hence we have the following scaling for 
various moments~\cite{REF:moments} of the joint probability distribution 
$P$ associated with localized particles ($\tilde{\ell}\ne 0$): 
\begin{mathletters}
\begin{eqnarray}
\int d\tau\,d\{\aaop\}\,{\cal D}\{\pacf\}\,
P(\tau,\{\aaop\},\{\pacf\})\,
\hat{\aaop}_{\hl\hm}
&\sim& 
\cp^{1+\tilde{\ell}/2},
\label{EQ:etamomscale}
\\
\int d\tau\,d\{\aaop\}\,{\cal D}\{\pacf\}\,
P(\tau,\{\aaop\},\{\pacf\})\,
\hat{\pacf}_{\hl\hm}(\hk) 
\exp\left(-\hk^{2}/2\tau\right)
&\sim& 
k^{\tilde{\ell}} 
\sim 
\cp^{\tilde{\ell}/2}, 
\end{eqnarray}%
\end{mathletters}%
the latter being valid for $\hk^{2}\sim\cp$.  Note the exponent 
${1+\tilde{\ell}/2}$ in Eq.~(\ref{EQ:etamomscale}). 

We now describe a plausible physical scenario that is consistent with
this scaling of $P$.  Near to the transition the network has many long
chains consisting of twice-bonded particles, with only the occasional
more highly-bonded particles linking them.  Most of the localized
particles are on extremely mobile segments.  Orbitals attached to
particles on such segments are likely to be even less localized
orientationally than orbitals on less mobile segments, such as those
near junctions between chains.  Consider, for the sake of illustration,
a dangling chain (i.e., one attached to the network only at one end).
The orientational fluctuations of successive orbitals compound the
fluctuations of orbitals further away from the junction, causing them to
be successively more delocalized.  This compounding of fluctuations so 
heavily suppresses orientational localization that it causes the 
scaling (and not just the numerical value) of the orientational 
localization characteristics $\aaop_{\ell m}$ to vary according to 
location in the network.  Moreover, because orientational localization
is heavily suppressed for such a large fraction of orbitals, the 
leading-order scaling of the moments of $\aaop_{\ell m}$ is dominated by 
contributions from the small fraction of better-localized orbitals 
(e.g., those near chain junctions), and is blind to the much larger 
fraction of less well-localized orbitals (e.g., those far from chain 
junctions).  This partitioning of localized orbitals into better 
and less well fractions yields a picture consistent with our results 
provided we assume that the better localized variety constitute a 
fraction of order $\cp$ of the localized orbitals.  This fraction 
manifests itself in the exponent ${1+\tilde{\ell}/2}$ in Eq.~(\ref{EQ:etamomscale}) as the additional $1$.  
This picture allows us to identify the following appealing scaling for 
the localization characteristics of the better localized orbitals: 
\begin{mathletters}
\begin{eqnarray}
\aaop_{\ell m}
&\sim&
\cp^{\ell /2}
\\
\pacf_{\ell m}(\hk)
&\sim& 
k^{\ell}\sim\cp^{\ell/2}, 
\end{eqnarray}%
\end{mathletters}%
the latter being valid for $\hk^{2}\sim\cp$. 
Whilst we cannot be certain that this scenario is a necessary 
consequence of our results, it is both consistent with them 
and physically plausible. 
\section{Concluding remarks}
\label{SEC:summary}
The primary result of this Paper is a statistical characterization of the
structure and heterogeneity of the equilibrium amorphous solid state
that emerges due to the random permanent covalent bonding of the
constituent particles.  This statistical characterization takes the form
of a joint probability distribution that ascertains the likelihood of
finding a particle:
(i)~to be localized, 
(ii)~to have a certain positional localization length, 
(iii)~for a bond connected to this particle to have a certain 
orientational localization characteristics, and 
(iv)~for the correlations between the thermal fluctuations in 
particle position and orbital orientation to have certain characteristics.

We expect the emerging statistical characterization to be valid beyond
the context of the model used to determine it.  The reason for this is
that, apart from certain simple dependences on physical parameters
describing the particles in the network, this characterization is in
fact a consequence of the order parameter that we have considered, the
symmetries of the  Landau-type free energy associated with this order
parameter, and a limited number of further assumptions. 

We have constructed an analytical approach to the equilibrium structure
of the amorphous solid state of a class of materials, such as silica
gels, formed by the permanent random covalent bonding of atoms or small
molecules.  However, the accuracy and scope of our results is limited in
the following significant ways.  We have focused on {\it equilibrium\/}
structural properties, we have worked {\it near\/} to the amorphous
solidification transition, and we have computed within the framework of
a {\it mean-field\/} approximation.  It would be interesting to have a
better understanding of the implications of these limitations, and to be
able to obtain results beyond them.

\noindent 
{\it Acknowledgments\/}: 
We thank Horacio Castillo, Reimer K\"uhn, Weiqun Peng and, 
especially, Annette Zippelius and Max Makeev for useful 
discussions.  This work was supported by the 
U.S.~National Science Foundation through grants 
DMR94-24511 (PMG) and 
GER93-54978 (KAS), and 
the Campus Research Board of the University of Illinois at 
Urbana-Champaign through grant 1-2-69454 (KAS). 
\appendix
\section{Some useful mathematical ingredients}
\label{APP:ingred}
Owing to the presence of the factor 
$\exp(i{\bf k}\cdot{\bf s})$ 
in the effective Hamiltonian, we shall 
need the Rayleigh expansion for plane waves:
\begin{equation}
\exp\left(i{\bf k}\cdot{\bf r}\right)=
4\pi\sum_{\ell=0}^{\infty}\sum_{m=-\ell}^{\ell}
i^{\ell}\,j_{\ell}(kr)\,
Y_{\ell m}^{\ast}({\bf k}/k)\,
Y_{\ell m}({\bf r}/r),
\label{EQ:rayleigh}
\end{equation}%
where $j_{\ell}(\rho)$ is a spherical Bessel function, i.e., 
$j_{\ell}(\rho)\equiv\sqrt{\pi/2\rho}\,J_{\ell+\half}(z)$. 
The small-$z$ asymptotics of $j_{\ell}(z)$ for small $\ell$ are:
\begin{mathletters} 
\begin{eqnarray} 
j_{0}(z)&\sim&1-z^{2}/6,\\
j_{1}(z)&\sim&z/3.
\end{eqnarray}%
\end{mathletters}%
We also make use of angular 
momentum one spherical components of vectors:
\begin{equation}
k_m^{\ast} \equiv \sqrt{{4 \pi \over {3}}}\, k\,Y^{\ast}_{1 m} ({{\bf k} / k}). 
\label{EQ:SCKdef}
\end{equation}
Lastly, we need some special values of Wigner $3-j$ symbols:
if $\ell_{1}+\ell_{2}+\ell_{3}$ is odd, then
\begin{equation}
 \left(	{{\ell}_{1}\atop{0}}
	{{\ell}_{2}\atop{0}}
	{{\ell}_{3}\atop{0}}\right) = 0,
\qquad
\left(	{1\atop{{m}_{1}}}
	{1\atop{{m}_{2}}}
	{0\atop{0}} \right) 
= \delta_{m_{1}+m_{2},0} (-1)^{m_{1}} \, {1\over{\sqrt{3}}}.
\end{equation}

\section{Averaging over the Deam-Edwards distribution}
\label{APP:DEDaverage}
In this Appendix, following the discussion in Ref.~\cite{REF:alloys}, we 
perform the disorder-averaging of the $n$-fold replicated partition 
function ${\tilde Z}^n$ with respect to the Deam-Edwards distribution ${\cal P}_{M}(\con)$
characterizing the distribution of the quenched random constraints 
$\con \equiv \big\{j_{e},j^{\prime}_{e};a_{e},a^{\prime}_{e}\big\}_{e=1}^{M}$ 
(i.e.,~bonds), defined in Sec.~\ref{SEC:DEdist}.
We start from the definition of the disorder average as a weighted average 
over all possible realizations of the disorder:
\begin{equation}
\big[{\tilde Z}^n(\con)\big] \equiv 
{\rm Tr}_{\con}\,{\cal P}_{M}(\con)\,{\tilde Z}^n(\con),
\end{equation}
where ${\rm Tr}_{\con}$ denotes a trace over all possible realizations of the
disorder $\con$, defined via:
\begin{equation}
{\rm Tr}_{\con}\cdots\equiv 
{1\over{\cal N}} \sum_{M=0}^{\infty}
\prod_{e=1}^{M}
\left(\,\sum_{j_e,a_e}\sum_{j'_e,a'_e}\,\right) \cdots.
\end{equation}
Here, $\cal N$ is a normalization factor, which we shall compute below, and
the sum over $M$ collects contributions from all possible numbers of
permanent covalent bonds. Note that our definition of the Deam-Edwards
distribution ${\cal P}_{M}(\con)$ includes the weighting factor
$(2 \pi \mu^2 V / N)^M / M!$ which is a quasi-Poissonian probability
distribution for the number of bonds $M$, controlled by the parameter $\mu^2$.
Hence, we have
\begin{equation}
\big[{\tilde Z}^n (\con)\big]
\equiv 
{1\over{\cal N}}\! \sum_{M=0}^{\infty}\!
{(2\pi V\mu^2/NA^{2})^M\over{M!}}\!
\prod_{e=1}^{M}\!\!
\left(\,\sum_{j_e,a_e}\sum_{j'_e,a'_e}\,\right) \!
\Big\langle
\prod_{e=1}^{M}
\delta^{(3)}({\bf c}_{j_{e}}+{1\over{2}}{\bf s}_{j_{e},a_{e}}-
{\bf c}_{j_{e}^{\prime}}-{1\over{2}}{\bf s}_{j_{e}^{\prime},a_{e}^{\prime}})\,
\Delta^{(2)}({\bf s}_{j_{e},a_{e}}+{\bf s}_{j_{e}^{\prime},a_{e}^{\prime}})
\Big\rangle_{N,1}^{(n+1)}.
\label{EQ:Zave}
\end{equation}
It is worth noting that the sums over $j_e$,$\,j'_e$,$\,a_e$, and $a'_e$ are
not restricted to preclude the situation where a particular orbital $a$ on
a particular particle $j$ is bonded to two (or more) other orbitals attached
to other particles, an unphysical situation, as covalent bonds are typically
regarded as ``saturable\rlap,'' i.e., any orbital can be bonded to at most one
other orbital. However, near the amorphous solidification transition the
bond density is low, thus we expect the occurrence of such multiply-bonded 
orbitals to be negligibly rare and thus not having much impact on the theory.

Introducing replicas, denoting by $\big\langle\cdots\big\rangle_{N,n+1}$ the
thermal average weighted by a Hamiltonian that does not couple the replicas,
and resumming the power series in $M$, we can rewrite Eq.~(\ref{EQ:Zave}) in
the final form
\begin{eqnarray}
\big[{\tilde Z}^n (\con)\big] 
&=& {\cal Z}_{n+1} / {\cal Z}_1, {\rm where}
\\
{\cal Z}_{n+1}
&\equiv&
\Bigg\langle
\exp\Bigg(
{2\pi V\mu^{2}
\over{NA^{2}}}
\sum_{j,j'=1}^{N}
\sum_{a,a'=1}^{A}
\prod_{\alpha=0}^{n}
\delta^{(3)}(
 {\bf c}_{j }^{\alpha}+\half{\bf s}_{j ,a }^{\alpha}
-{\bf c}_{j'}^{\alpha}-\half{\bf s}_{j',a'}^{\alpha})\,
\Delta^{(2)}(
 {\bf s}_{j ,a }^{\alpha}, 
-{\bf s}_{j',a'}^{\alpha})
\Bigg)\Bigg\rangle_{N,n+1}.
\end{eqnarray}

\section{Denominator for the Self-Consistent Equation}
\label{APP:denominator}
In this Appendix we evaluate the denominator of the right-hand side
of the SCE for $\fop(\hk;\hl,\hm)$, Eq.~(\ref{EQ:SCEc}), namely, the quantity
\begin{equation}
{\displaystyle
\left\langle\exp\bigg(
{4\pi\mu^{2}\over {V^{n}}}
\sum_{\hat{\bf k}\hat{\ell}\hat{m}}
\pop(\hk;\hl,\hm)\,
{\rm e}^{i\hat{\bf k}\cdot\hat{\bf c}}\,
{1\over{A}}\sum_{a=1}^{A}
{\rm e}^{i\hat{\bf k}\cdot\hs_{a}}
\hy_{\hl \hm}(-\hs_a)
\bigg)\right\rangle_{1,n+1} }.
\label{EQ:denom1}
\end{equation}
To do this, we expand the exponential in a power series and consider 
the $r$-th order term:
\begin{eqnarray}
&&
\left\langle 
{1\over{r !}}
\left( 4 \pi \mu^{2} \right)^{r}  q^r V^{-nr}
\sum_{\hk_1 \hl_1 \hm_1}
\sum_{\hk_2 \hl_2 \hm_2}
\ldots
\sum_{\hk_r \hl_r \hm_r}
\pop(\hk_1;\hl_1,\hm_1)\,
\pop(\hk_2;\hl_2,\hm_2)\,
\ldots
\pop(\hk_r;\hl_r,\hm_r)\,
{\rm e}^{i\hat{\bf k}_1\cdot\hat{\bf c}}\,
{\rm e}^{i\hat{\bf k}_2\cdot\hat{\bf c}}\,
\ldots
{\rm e}^{i\hat{\bf k}_r\cdot\hat{\bf c}}\,
\right.
\nonumber
\\
&&\qquad\times \left.
{1\over{A^r}}\sum_{a_1, \ldots, a_r}
{\rm e}^{i\hat{\bf k}_1\cdot\hs_{{a}_1}}
{\rm e}^{i\hat{\bf k}_2\cdot\hs_{{a}_2}}
\ldots
{\rm e}^{i\hat{\bf k}_r\cdot\hs_{{a}_r}}
\hy_{\hl_1 \hm_1}(-\hs_{a_1})
\ldots
\hy_{\hl_r \hm_r}(-\hs_{a_r})
\right\rangle_{1,n+1} .
\label{EQ:denom2}
\end{eqnarray}%
Noticing that the quantity (\ref{EQ:denom2}) factorizes 
over the replicas, and passing to the replica limit $n \to 0$, we rewrite
\begin{eqnarray}
&&
\left\langle {1\over{r !}}
\left(4\pi\mu^{2}\right)^{r} q^r
\sum_{{\bf k}_1 \ell_1 m_1}
\sum_{{\bf k}_2 \ell_2 m_2}
\ldots
\sum_{{\bf k}_r \ell_r m_r}
\pop({\bf k}_1; \ell_1, m_1)\,
\pop({\bf k}_2; \ell_2, m_2)\,
\ldots
\pop({\bf k}_r; \ell_r, m_r)\,
{\rm e}^{i {\bf k}_1\cdot {\bf c}}\,
{\rm e}^{i {\bf k}_2\cdot {\bf c}}\,
\ldots
{\rm e}^{i {\bf k}_r\cdot {\bf c}}\,
\right.
\nonumber
\\
&&\qquad\times\left.
{1\over{A^r}} \sum_{a_1, \ldots, a_r}
{\rm e}^{i {\bf k}_1\cdot {\bf s}_{{a}_1}}
{\rm e}^{i {\bf k}_2\cdot {\bf s}_{{a}_2}}
\ldots
{\rm e}^{i {\bf k}_r\cdot {\bf s}_{{a}_r}}
Y_{\ell_1 m_1 }(-{\bf s}_{a_1})
\ldots
Y_{\ell_r m_r}(-{\bf s}_{a_r})
\right\rangle_{1,1} .
\label{EQ:denom3}
\end{eqnarray}%
By the MTI of $\pop(\hk; \hl, \hm)$ we know that $\tilde{\bf k}_i = {\bf 0}$, 
which means that ${\bf k}_i = {\bf 0}$.  Also, using MRI of $\pop(\hk; \hl, \hm)$
we see that if $\hk = \bfhz$ then (\ref{EQ:denom3}) can only be nonzero if
$\hl=\hz$ also. Knowing that $\pop(\bfhz; \hz, \hz)=1/\sqrt{4 \pi}$, we
get, for the $r$-th order contribution ${(\mu^2)}^r q^r /r!$.
Resumming the power series we obtain, finally
\begin{equation}
{\displaystyle
\left\langle\exp\bigg(
{4\pi\mu^{2} \over {V^{n}}}
\sum_{\hat{\bf k}\hat{\ell}\hat{m}}
\pop(\hk;\hl,\hm)\,
{\rm e}^{i\hat{\bf k}\cdot\hat{\bf c}}\,
{1\over{A}}\sum_{a=1}^{A}
{\rm e}^{i\hat{\bf k}\cdot\hat{\bf s}_{a}}
\hy_{\hl \hm}(\hs_{a})
\bigg)\right\rangle_{1,n+1} }
= \sum_{r=0}^{\infty} {{{(\mu^2)}^r q^r}\over{r !}} 
= \exp(\mu^2 q).
\label{EQ:denomf}
\end{equation}

\section{Definitions of the $\bfpc{n}$ Constants}
\label{APP:BFPC}
The $\bfpc{n}$ constants are defined as follows:
\begin{mathletters}
\begin{eqnarray}
\bfpc{0}&\equiv&{A^{-1}\left(1+(A-1)\co{1}\right)}
\\
\bfpc{1}&\equiv&{1\over{1+\bfpc{0}}}
\\
\bfpc{2}&\equiv&{{1\over{A}}\left(1+(A-1)\co{1}^{2}\right)}
\\
\bfpc{3}&\equiv&{\bfpc{2} - 2 \bfpc{2} \bfpc{1}}
\\
\bfpc{4}&\equiv&{{1-\bfpc{2}\over{\bfpc{2}}}}
\\
\bfpc{5}&\equiv&{{\bfpc{3}\over{3 \bfpc{2} \bfpc{4}}}}
\\
\bfpc{6}&\equiv&{{\bfpc{3}\over{12 \bfpc{2} \bfpc{4}^2 }} +
{\bfpc{1}^2\over{8 \bfpc{4}}} }
\end{eqnarray}%
\end{mathletters}%
They encode information about the strength of the mutual repulsion of orbitals.
\section{Anisotropic self--consistent equation: 
angular momentum one in one replica channel}
\label{APP:ASCE0100}
In this Appendix we study in detail the \sce\ for the $\pop\vtripW{1}(\hk)$ 
component. We start with the full form:

\begin{eqnarray}
q \pop\vtripW{1}(\hk) =&\phantom{+}& (1-q)
{4\pi\mu^{2} \over {V^{n}}}\,q
\sum_{\hk_1,\hl_{1},\hm_{1}}
\pop(\hk_{1};\hl_{1},\hm_{1})\,
\nonumber
\\
&&\qquad
\times{1\over{A^{2}}}\!
\sum_{a_{0},a_{1}=1}^{A}\!
\left\langle
{\rm e}^{-i\hk \cdot\hat{\bf c}}\!
\prod_{\alpha\ne\alpha_1}\!
\left(Y_{0 0} ({\bf s}_{a_{0}}^{\alpha})\right)\,
Y_{1 m_1}^{\ast} ({\bf s}_{a_{0}}^{\alpha_1})\,
{\rm e}^{i\hk_{1}\cdot\hat{\bf c}}\,\,
{\rm e}^{i\hk_{1}\cdot\hat{\bf s}_{a_{1}}}\,
\hy_{\hl \hm}(-\hs_{a_1})
\right\rangle_{1,n+1}
\nonumber
\\
&+&\frac{1}{2}
\left({4\pi\mu^{2} \over {V^{n}}}\right)^{2}\,q^{2}
\sum_{\hk_1,\hl_{1},\hm_{1}}
\sum_{\hk_2,\hl_{2},\hm_{2}}
\pop(\hk_{1};\hl_{1},\hm_{1})\,
\pop(\hk_{2};\hl_{2},\hm_{2})
\nonumber
\\
&&\qquad\times
{1\over{A^{3}}}\sum_{a_{0},a_{1},a_{2}=1}^{A}
\Bigg\langle
{\rm e}^{-i\hk\cdot\hat{\bf c}}
\prod_{\alpha\ne\alpha_1}^{n}
\left(Y_{0 0} ({\bf s}_{a_{0}}^{\alpha_{0}})\right)\,
Y_{1 m_1}^{\ast} ({\bf s}_{a_{0}}^{\alpha_1})
{\rm e}^{i\hk_{1}\cdot\hat{\bf c}}\,
{\rm e}^{i\hk_{1}\cdot\hat{\bf s}_{a_{1}}}
\nonumber
\\
&&\qquad\qquad\times
\hy_{\hl \hm}(\hs_{a_1})
{\rm e}^{i\hk_{2}\cdot\hat{\bf c}}\,
{\rm e}^{i\hk_{2}\cdot\hat{\bf s}_{a_{2}}}
\hy_{\hl \hm}(\hs_{a_2})
\Bigg\rangle_{1,n+1}
\label{EQ:A0100G}
\end{eqnarray}%
We now proceed to study this equation order by order, starting with 
the linear term. As we are not interested in feedback on this component 
from higher angular momentum components we may truncate the angular momentum
sums to include only terms of ${\hl^2} \le 2$. Thus we must include couplings
to the components $\pop\vtripW{2}(\hk)$ and $\pop(\hk)$ only.  We study the coupling
to $\pop(\hk)$ first (i.e.,~$\hl_1=\hm_1=0$). Clearly then in replica $\alpha_1$ 
we must expand the shift-factor 
$\exp(i {\bf k}_{1}^{\alpha}\cdot{\bf s}^{\alpha}_{a_{1}})$ 
to angular momentum 1 and put this factor equal to unity in all other replicas,
obtaining

\begin{eqnarray}
&&(1-q) {4\pi\mu^{2} \over {V^{n}}}\,q
\sum_{\hk_1}
\pop(\hk_{1})\,
\nonumber
\\
&&\qquad
\times{1\over{A^{2}}}
\sum_{a_{0},a_{1}=1}^{A}
\left\langle
{\rm e}^{-i\hk \cdot\hat{\bf c}}
{\rm e}^{i\hk_{1}\cdot\hat{\bf c}}\,
\left( {1\over{\sqrt{4 \pi}}} \right)^n
Y_{1 m_1}^{\ast} ({\bf s}_{a_{0}}^{\alpha_1})
\left( {1\over{\sqrt{4 \pi}}} \right)^{(n+1)}
4 \pi i \,\, j_1 (k_1^{\alpha_1}) \, \sum_{m'=-1}^{1}
Y_{1 m'} ({\bf s}_{a_{1}}^{\alpha_1})\,
Y_{1 m'}^{\ast} \left( {{{{\bf k}_1}^{\alpha_1}}\over{k_1^{\alpha_1}}}\right)
\right\rangle_{1,n+1}
\nonumber
\\
&&\qquad
= (1-q) \mu^2 i \, {\sqrt{4 \pi} \over {3}} \pop(\hk) 
{1\over{A^{2}}} \sum_{a_{0},a_{1}=1}^{A}
\Big(\delta_{a,a_1}+(1-\delta_{a,a_1})\,\co{1}\Big) 
k^{\alpha_1} \, 
Y_{1 m_1}^{\ast} 
\left( { {\bf k} ^ {\alpha_1} \over {k^{\alpha_1} }} \right)
\nonumber
\\
&&\qquad
= {i\over{\sqrt{3}}}\, \bfpc{0} \pop(\hk)\,\sck{1}.
\label{EQ:A01001B}
\end{eqnarray}%

We now study the linear term corresponding to the coupling of 
$\pop\vtripW{1}(\hk)$ to $\pop\vtripW{2}(\hk)$. When $\alpha_1 \ne \alpha_2$
this term, being of order $\hk^2$, is subdominant. Thus we need
only consider the term in the angular momentum sum where $\ell^{\alpha_1} = 1$,
$\hl_1$ being $0$ in all other replicas. We examine this term in the sum, 
making use of the value of the two-orbital correlator, given by 
Eq.~(\ref{EQ:BareCor}):
\begin{eqnarray}
&&(1-q) {4\pi\mu^{2} \over {V^{n}}}\,q
\sum_{\hk_1}
\pop\vtripW{1}(\hk_1)\,
{1\over{A^{2}}}
\sum_{a_{0},a_{1}=1}^{A}
\left\langle
{\rm e}^{-i\hk \cdot\hat{\bf c}}\,\,
{\rm e}^{i\hk_{1}\cdot\hat{\bf c}}\,
\left( {1\over{\sqrt{4 \pi}}} \right)^n
Y_{1 m_1}^{\ast} ({\bf s}_{a_{0}}^{\alpha_1})
\left( {1\over{\sqrt{4 \pi}}} \right)^n
Y_{1 m_1} (-{\bf s}_{a_{1}}^{\alpha_1})
\right\rangle_{1,n+1}
\nonumber
\\
&&\qquad
= - (1-q) \mu^2 \pop\vtripW{1}(\hk) {1\over{A^{2}}} \sum_{a_{0},a_{1}=1}^{A}
 \Big(\delta_{a,a_1}+(1-\delta_{a,a_1})\,\co{1}\Big)
\nonumber
\\
&&\qquad
= - (1-q)\, \bfpc{0}\, \mu^2 \pop\vtripW{1}(\hk).
\label{EQ:A01001A}
\end{eqnarray}%

We next proceed to examine the second-order term in Eq.~(\ref{EQ:A0100G}). 
Again, we truncate the
angular momentum sums. As we are only interested in the leading-order contributions
to $\pop\vtripW{1}(\hk)$ there are only two cases to consider: $\hl_1=\hl_2=\hz$
and $\hl_1^2=2,\, \hl_2^2 = 0$. The former case, however, results in a term that is
subdominant to term (\ref{EQ:A01001B}). We thus consider the latter case:
\begin{eqnarray}
{1\over{2}} &&(1-q) \left( {4\pi\mu^{2} \over {V^{n}}} \right)^2 q^2
\sum_{\hk_1, \hk_2}
\pop(\hk_1)\,\pop\vtripW{1}(\hk_2)
\nonumber
\\
&&\qquad
{1\over{A^{2}}}
\sum_{a_{0},a_{1}=1}^{A}
\left\langle
{\rm e}^{-i\hk \cdot\hat{\bf c}}
{\rm e}^{i\hk_{1}\cdot\hat{\bf c}}\,
{\rm e}^{i\hk_{2}\cdot\hat{\bf c}}\,
\left( {1\over{\sqrt{4 \pi}}} \right)^{(n+1)}
\left( {1\over{\sqrt{4 \pi}}} \right)^n
Y_{1 m_1}^{\ast} ({\bf s}_{a_{0}}^{\alpha_1})
\left( {1\over{\sqrt{4 \pi}}} \right)^n
Y_{1 m_1} (-{\bf s}_{a_{1}}^{\alpha_1})
\right\rangle_{1,n+1}
\nonumber
\\
&&\qquad
= - {\sqrt{4 \pi}\over{2}} (1-q) {\mu^4 \over {V^{n}}} 
\sum_{\hk_1} q^2 \pop(\hk_1)\,\pop\vtripW{1}(\hk - \hk_1)
{1\over{A^{2}}} \sum_{a_{0},a_{1}=1}^{A}
 \Big(\delta_{a,a_1}+(1-\delta_{a,a_1})\,\co{1}\Big)
\nonumber
\\
&&\qquad
= -{\sqrt{4 \pi}\over{2}} (1-q) \bfpc{0}\, 
{\mu^4 \over {V^{n}}} \sum_{\hk_1} q^2 \pop(\hk_1)\,\pop\vtripW{1}(\hk - \hk_1).
\label{EQ:A01002D}
\end{eqnarray}%
Keeping in mind that due to the symmetry $\hl_1 \leftrightarrow \hl_2$ we must
double the term~(\ref{EQ:A01002D}), we now assemble all the terms, dropping
sub-leading contributions:
\begin{equation}
q\pop\vtripW{1}(\hk)=
	\bfpc{0}\bigg(
 - q\pop\vtripW{1}(\hk)
+{i\over{\sqrt{3}}}
 q\pop(\hk_{1})\,\sck{1}
-q^2{\sqrt{4\pi}\over{V^{n}}}\sum\nolimits_{\hk_{1}}
 \pop(\hk-\hk_{1})\,
 \pop\vtripW{1}(\hk_{1})
	\bigg).
\end{equation}
Cancelling a factor of $q$ leaves us with Eq.~(\ref{EQ:WRAsce}).

\section{Anisotropic self--consistent equation: 
angular momentum one in two replica channels}
\label{APP:ASCE0110}
In this Appendix we study in detail the \sce\ for the 
$\pop\vtripW{1}\vtripW{2}(\hk)$ 
component. We start with the full form:
\begin{eqnarray}
q \pop\vtripW{1}\vtripW{2}(\hk) &=& (1-q)
{4\pi\mu^{2} \over {V^{n}}}\,q
\sum_{\hk_1,\hl_{1},\hm_{1}}
\pop(\hk_{1};\hl_{1},\hm_{1})\,
\nonumber
\\
&&\qquad
{1\over{A^{2}}}
\sum_{a_{0},a_{1}=1}^{A}
\left\langle
{\rm e}^{-i\hk \cdot\hat{\bf c}}
\prod_{\alpha\ne\alpha_1,\alpha_2}
\left(Y_{0 0} ({\bf s}_{a_{0}}^{\alpha})\right)\,
Y_{1 m_1}^{\ast} ({\bf s}_{a_{0}}^{\alpha_1})
Y_{1 m_2}^{\ast} ({\bf s}_{a_{0}}^{\alpha_2})
{\rm e}^{i\hk_{1}\cdot\hat{\bf c}}\,
{\rm e}^{i\hk_{1}\cdot\hat{\bf s}_{a_{1}}}
\hy_{\hl \hm}(\-\hs_{a_1})
\right\rangle_{1,n+1}
\nonumber
\\
&&
+\frac{1}{2}
\left({4\pi\mu^{2} \over {V^{n}}}\right)^{2}\,q^{2}
\sum_{\hk_1,\hl_{1},\hm_{1}}
\sum_{\hk_2,\hl_{2},\hm_{2}}
\pop(\hk_{1};\hl_{1},\hm_{1})\,
\pop(\hk_{2};\hl_{2},\hm_{2})
\nonumber
\\
&&\qquad\times
{1\over{A^{3}}}\sum_{a_{0},a_{1},a_{2}=1}^{A}
\Bigg\langle
{\rm e}^{-i\hk\cdot\hat{\bf c}}
\prod_{\alpha\ne\alpha_1,\alpha_2}
Y_{0 0} ({\bf s}_{a_{0}}^{\alpha_{0}})
Y_{1 m_1}^{\ast} ({\bf s}_{a_{0}}^{\alpha_1})
Y_{1 m_2}^{\ast} ({\bf s}_{a_{0}}^{\alpha_2})
{\rm e}^{i\hk_{1}\cdot\hat{\bf c}}\,
{\rm e}^{i\hk_{1}\cdot\hat{\bf s}_{a_{1}}}
\nonumber
\\
&&\qquad\qquad\times
\hy_{\hl \hm}(\-\hs_{a_1})
{\rm e}^{i\hk_{2}\cdot\hat{\bf c}}\,
{\rm e}^{i\hk_{2}\cdot\hat{\bf s}_{a_{2}}}
\hy_{\hl \hm}(\-\hs_{a_2})
\Bigg\rangle_{1,n+1}
\label{EQ:A0110G}
\end{eqnarray}%
As we did before in App.~\ref{APP:ASCE0100} for the \sce\ for 
$\pop\vtripW{1}(\hk)$ we proceed to study the linear terms of this
equation. Again we truncate to include only the components of 
$\pop$ of angular momentum smaller than the angular momentum of 
$\pop\vtripW{1}\vtripW{2}(\hk)$, i.e. $\ell^\alpha = 0$ for
$\alpha \ne \alpha_1, \alpha_2$ and $\ell^{\alpha_1}, \ell^{\alpha_2} \le 1$. 
Thus we find couplings to terms $\pop(\hk)$, $\pop\vtripW{1}(\hk)$,
$\pop\vtripW{2}(\hk)$, and $\pop\vtripW{1}\vtripW{2}(\hk)$. Expanding
the shift-factor 
$\exp({i {\bf k}_{1}^{\alpha}\cdot{\bf s}^{\alpha}_{a_{1}}})$
to angular momentum $0$ or $1$ as appropriate in each replica, proceeding
as in Eqs. (\ref{EQ:A01001A}) and (\ref{EQ:A01001B}), and, finally, 
inserting the value for $\pop\vtripW{1}(\hk)$ that we obtained in 
Sec.~\ref{SEC:aniso0100},
we obtain the linear contribution (terms are in the order listed above):
\begin{eqnarray}
&&
(1-q) {4\pi\mu^{2}\over{ V^{n}}}\,q
\sum_{\hk_1}
{1\over{A^{2}}}\sum_{a_{0},a_{1}=1}^{A}
\Bigg\langle
{\rm e}^{-i\hk \cdot\hat{\bf c}}
{\rm e}^{i\hk_{1}\cdot\hat{\bf c}}\,
\left( {1\over{\sqrt{4 \pi}}} \right)^{n-1}
Y_{1 m_1}^{\ast} ({\bf s}_{a_{0}}^{\alpha_1})
Y_{1 m_2}^{\ast} ({\bf s}_{a_{0}}^{\alpha_2})
\times
\nonumber
\\
&&\qquad
\Bigg\{ \pop(\hk_{1})\,
\left( {1\over{\sqrt{4 \pi}}} \right)^{n+1}
(4 \pi i)^2\, j_1 (k_1^{\alpha_1}) \,j_1 (k_1^{\alpha_2}) \,
\sum_{m'_1,m'_2=-1}^{1}
Y_{1 m'_1} ({\bf s}_{a_{1}}^{\alpha_1})\,
Y_{1 m'_1}^{\ast} \left( {{{{\bf k}_1}^{\alpha_1}}\over{k_1^{\alpha_1}}}\right)\,
Y_{1 m'_2} ({\bf s}_{a_{1}}^{\alpha_2})\,
Y_{1 m'_2}^{\ast} \left( {{{{\bf k}_2}^{\alpha_2}}\over{k_1^{\alpha_2}}}\right)
\nonumber
\\
&&\qquad\qquad
+\pop\vtripW{1}(\hk_1)\,
\left( {1\over{\sqrt{4 \pi}}} \right)^n
Y_{1 m_1} (-{\bf s}_{a_{1}}^{\alpha_1})
4 \pi i \,\, j_1 (k_1^{\alpha_2}) \, \sum_{m'=-1}^{1}
Y_{1 m'} ({\bf s}_{a_{1}}^{\alpha_2})\,
Y_{1 m'}^{\ast} \left( {{{{\bf k}_1}^{\alpha_2}}\over{k_1^{\alpha_2}}}\right)
\nonumber
\\
&&\qquad\qquad
+\pop\vtripW{2} (\hk_1)\,
\left( {1\over{\sqrt{4 \pi}}} \right)^n
Y_{1 m_2} (-{\bf s}_{a_{1}}^{\alpha_2})
4 \pi i \,\, j_1 (k_1^{\alpha_1}) \, \sum_{m'=-1}^{1}
Y_{1 m'} ({\bf s}_{a_{1}}^{\alpha_1})\,
Y_{1 m'}^{\ast} \left( {{{{\bf k}_1}^{\alpha_1}}\over{k_1^{\alpha_1}}}\right)
\nonumber
\\
&&\qquad\qquad
+\pop\vtripW{1}\vtripW{2} (\hk_1)\,
\left( {1\over{\sqrt{4 \pi}}} \right)^{n-1}
Y_{1 m_1} (-{\bf s}_{a_{1}}^{\alpha_1})
Y_{1 m_2} (-{\bf s}_{a_{1}}^{\alpha_2})
\Bigg\}
\Bigg\rangle_{1,n+1}
\nonumber
\\
&&
= q\, \bfpc{2} \left(
- {1\over{3}}\sck{1}\,\sck{2}\,\pop(\hk)
- {i\over{\sqrt{3}}}\sck{2} \pop\vtripW{1}(\hk)
- {i\over{\sqrt{3}}}\sck{1} \pop\vtripW{2}(\hk)
+ \pop\vtripW{1}\vtripW{2}(\hk)
\right)
\nonumber
\\
&&
= \bfpc{2}q\,\pop\vtripW{1}\vtripW{2}(\hk)
-{q\over{3}}\bfpc{3}
 \sck{1}\,\sck{2}\,\pop(\hk).
\label{EQ:A01101}
\end{eqnarray}%

We next proceed to study the quadratic contribution to 
$\pop\vtripW{1}\vtripW{2}(\hk)$ from Eq.~(\ref{EQ:A0110G}). Truncating the
angular momentum sums as usual we see that there are the following four cases 
to consider:

(i)~$\hl_1=\hl_2=0$, corresponding to the coupling to $\pop(\hk_1)\pop(\hk_2)$

(ii)~$\hl_1=0, \hl_2^2=2$, corresponding to the coupling to
$\pop(\hk_1)\pop\vtripW{1}(\hk_2)$

(iii)~$\hl_1^2=\hl_2^2=2$, corresponding to the coupling to
$\pop\vtripW{1}(\hk_1)\pop\vtripW{2}(\hk_2)$

(iv)~$\hl_1=0, \hl_2^2=4$  corresponding to the coupling to
$\pop(\hk_1)\pop\vtripW{1}\vtripW{2}(\hk_2)$

\noindent
We study these terms in more detail, as in Appendix~\ref{APP:ASCE0100},
Eq.~(\ref{EQ:A01002D}). Term (i) is clearly subdominant compared with the 
second term in Eq.~(\ref{EQ:A01101}). Writing out the remaining terms:
\begin{eqnarray}
&&{\rm term (ii):} -{i\over{\sqrt{3}}} {\sqrt{4 \pi}\over{2}} (1-q) \bfpc{2}\, 
{\mu^4 \over {V^{n}}} \sum_{\hk_1} q^2 \sck{2} 
\pop(\hk_1)\,\pop\vtripW{1}(\hk - \hk_1),
\\
&&{\rm term (iii):} {\sqrt{4 \pi}\over{2}} (1-q) \bfpc{2}\, 
{\mu^4 \over {V^{n}}} \sum\nolimits_{\hk_1} q^2
\pop\vtripW{1}(\hk_1)\,\pop\vtripW{2}(\hk - \hk_1),
\\
&&{\rm term (iv):} {\sqrt{4 \pi}\over{2}} (1-q) \mu^4 \bfpc{2}\, 
{\mu^4 \over {V^{n}}} \sum\nolimits_{\hk_1} q^2
\pop(\hk_1)\,\pop\vtripW{1}\vtripW{2}(\hk - \hk_1).
\end{eqnarray}
Term (ii) does not contribute to the leading-order behavior of 
$\pop\vtripW{1}\vtripW{2}(\hk)$ as for $\hk^2$ of order $\cp$ or greater all
quadratic terms are subleading compared to linear terms, 
and for  $\hk^2 \ll \cp$  it will, because
of the $\sck{2}$ factor, be subdominant compared to terms (iii)
and (iv). 
Keeping in mind that because of the symmetry  $\hl_1 \leftrightarrow \hl_2$
we must double the term (iv), we reassemble the pieces of the \sce\ for
$\pop\vtripW{1}\vtripW{2}(\hk)$, dropping subleading contributions:

\begin{eqnarray}
q\pop\vtripW{1}\vtripW{2}(\hk)
&=&
 \bfpc{2} q \pop\vtripW{1}\vtripW{2}(\hk)
-{1\over{3}} \bfpc{3}
 \sck{1}\,\sck{2}\, q \pop(\hk)
+q^2 \bfpc{2} {\sqrt{4\pi} \over {V^{n}}} \sum_{\hk_{1}}
 \pop(\hk-\hk_{1})\,\pop\vtripW{1}\vtripW{2}(\hk_{1})
\nonumber
\\
&&\qquad\qquad
+q^2 \bfpc{2} {\sqrt{4\pi}\over{2V^{n}}} \sum_{\hk_{1}} 
 \pop\vtripW{1}(\hk-\hk_{1})\,\pop\vtripW{2}(\hk_{1}),
\end{eqnarray}%
Cancelling a factor of $q$ leaves us with Eq.~(\ref{EQ:TRAsce}).



\begin{references}
\bibitem{REF:qualif}
What we, in fact, compute are moments of averages of such distributions 
over realizations of the quenched randomness.  Although we have 
not checked this point in detail, we do not expect there 
to be sample-to-sample variations of the distribution.
\bibitem{REF:stability}
As we are not able to compute the free energy for arbitrary values of
the order parameter, we are unable to rule out the possibility, that a
discontinuous phase transition occurs at a bond density smaller than
$\mu_c^2$. That the transition is continuous should be regarded as an
additional assumption of ours.
\end{references}
\end{document}